\documentclass[a4paper,11pt]{article}
\pdfoutput=1 

\usepackage{jcappub} 

\usepackage[T1]{fontenc} 
\usepackage{subcaption}
\usepackage{caption}
\usepackage{dsfont}
\usepackage{physics}
\usepackage{amssymb}
\usepackage{hyperref}
\usepackage{ulem}

\title{\boldmath \texttt{Pz Cats}: Photometric redshift catalogs based on DES Y3 BAO sample}

\author[a]{Paula S. Ferreira,}
\author[a,b]{Ribamar R. R. Reis}
\affiliation[a]{Instituto de Física, Universidade Federal do Rio de Janeiro,\\
Av. Athos da Silveira Ramos, 149 - Cidade Universitária, Rio de Janeiro, Brasil}

\affiliation[b]{Observatório do Valongo, Universidade Federal do Rio de Janeiro\\
Ladeira Pedro Antônio, 43, Centro,  Rio de Janeiro,
Brasil}
\emailAdd{psfer@pos.if.ufrj.br, ribamar@if.ufrj.br}

\abstract{Over the years, photometric redshift estimation (photo-z) has advanced through various methods. This study evaluates four distinct photo-z estimators—\texttt{ANNz2}, \texttt{BPZ}, \texttt{ENF}, and \texttt{DNF}—using the Dark Energy Survey Y3 BAO Sample. Unlike most studies, we explore selecting optimal galaxies based on their redshift Probability Distribution Function (PDF) by either reducing noise or identifying those approximating a Gaussian distribution. We cross-matched 25,760 galaxies drawn from four spectroscopic surveys with the photo-z sample to comprehend redshift bias and its 68th percentile $\sigma_{68}$. The lowest $\sigma$ for all estimators was found in the range $0.79<z_p<0.85$. Among the estimators, \texttt{DNF} exhibited the greatest bias, while \texttt{ENF}, \texttt{ANNz2}, and \texttt{BPZ} showed decreased precision outside 0.7 to 0.9 redshift range. To select galaxies with minimal bias, \texttt{ANNz2} emerged as the most reliable algorithm across all criteria. PDFs selection significantly improves colour representation over the spectroscopic sample, underscoring the role of magnitude space in selection. While \texttt{ANNz2} achieved superior precision, \texttt{ENF} poorly selected Gaussian PDFs, leaving few galaxies for LSS evaluation. Despite smooth PDFs, catastrophic redshift errors were present. Though \texttt{DNF} had the poorest precision, it offered enough galaxies for cosmological use. Subsampling galaxies with secondary peaks less than 30\% of the main peak height, termed Small Peaks, showed \texttt{ANNz2} excelled. The catalogs produced have been published as \texttt{Pz Cats} within the \texttt{ZENODO} repository.}


\begin{document}
\maketitle
\flushbottom

\section{Introduction}

Redshift estimation can be obtained in two ways via spectroscopy or photometry. Spectroscopic surveys rely on spectrographs and optical fibres, and this requires a long exposure time of a single object to later compare the spectrum to a measured spectrum in a laboratory. Photometric surveys are based on bands/filters in different wavelength ranges. For this second type of survey construction, we need to estimate the redshift of each observed object from the measured magnitudes and compare them to a set of previously observed objects whose spectroscopy and photometry are known.

Photometric surveys are ideal for observing a larger
number of objects, rather than spectroscopic ones, and understand the properties of their shape to infer their morphology. The methods used can be template fitting or training-based machine learning methods. Template fitting is based on the inference of photometric redshift (photo-z) that requires a set of known objects that should include all their sub-types based on empirical and/or theoretical models which allow us to predict the photometry from a few parameters, like stellar mass, star formation rate and redshift, among others. Some examples of algorithms are: \texttt{LePhare} \cite{arnouts1999measuring,guzzo2014vimos}, \texttt{BPZ} \cite{benitez2000bayesian}, \texttt{ZEBRA}\cite{feldmann2006zurich}, \texttt{EAZY} \cite{brammer2008eazy}. Training-based methods require a set of coincident galaxies whose spectroscopic redshifts are known, some examples of algorithms are: \texttt{TPZ} \cite{10.1093/mnras/stt574}, \texttt{ANNz2} \cite{sadeh2016annz2}, \texttt{GPZ} \cite{almosallam2016gpz}, \texttt{DNF}\cite{de2016dnf}.

Luminous Red Galaxies (LRG) are the best sources for surveys whose goal is to study the Large Scale Structure (LSS). They are correlated with clusters, occupying massive halos \cite{padmanabhan2005calibrating,hoshino2015luminous}. LRGs are bright and red, they have uniform Spectral Energy Distribution (SED) which makes them ideal for photo-z surveys \cite{eisenstein2001spectroscopic}. In this study, we will focus on the Dark Energy Survey (DES) LRGs for the BAO sample. They used the VISTA Hemisphere Survey (VHS) \cite{mcmahon2012vista} together with DES filters to select the LRGs in the $rzK$-space \cite{banerji2015combining} for redshifts above $0.5$.

There are many thorough photometric redshift verifications made by the DES Collaboration or independent groups like \cite{sevilla2021dark,sanchez2014photometric,san2024dark}. Our main goal in this paper is to analyse how selecting galaxies through their full redshift Probability Distribution Function (PDF) can affect the effective redshift ($z_{\rm eff}$) of the survey, its precision and sample size. This is different than assuming the PDFs are nearly Gaussian and comparing different photo-z estimators by their redshift variance. Furthermore, we are using a different training spectroscopic sample, with intersections with the ones used by the DES Collaboration, less populated, but still relevant to understand how each estimator can be influenced by the size of the training sample.  In terms of precision, we obtain the photo-z errors without assuming any distribution function to the sample's distribution $N(z)$.

The present study is organized into five sections. Section \ref{sec:estimators} describes the photometric estimators we used for this analysis. Section \ref{sec:bias_bin} describes the error estimation of the photometric methods. Next, we show in section \ref{sec:results} the results for each estimator and their respective cuts, including selecting the best redshift distribution functions and colour cut. The last section \ref{sec:conclusion}
summarises the study.

\section{Photo-z estimators}\label{sec:estimators}
In this work, we used the Dark Energy Survey BAO sample of red galaxies from the third year of observations, Dark Energy Survey (DES) Year 3 \cite{abbott2022dark} photometric sample. The dataset collected from the survey comprises a total of $7,081,993$ galaxies, spread over an area of $21,078.79$ square degrees. This amounts to an observed density of approximately $333.60$ galaxies per square degree. The input of each object was used in the three photo-z estimators within a range of $0<z<2.0$. We test four different algorithms for estimating the Probability Density Function: \texttt{ANNz2}, \texttt{BPZ}, \texttt{ENF}, \texttt{DNF} \footnote{We included the codes used here at: \url{https://github.com/psilvaf/cat_org}.}.
\subsection{BPZ}

The Bayesian photometric redshift estimation (BPZ) \cite{benitez2000bayesian} version \texttt{1.98b} is a Bayesian probability redshift estimation algorithm, a template fitting method. This method is based on the Bayesian theory that all probabilities are conditional. Thus, the redshift of a galaxy given that we have photometric information will be $p(z|\mathbf{m_0})$, where $z$ is the redshift we want to find and $m_0$ are the magnitudes in each filter. 

Applying Bayes' theorem, we have
\begin{equation}
    p(z|\mathbf{m_0})= \frac{p(\mathbf{m_0}|z)p(z)}{p(\mathbf{m_0})},
\end{equation}
where $p(\mathbf{m_0}|z)$ is the likelihood, the probability that the magnitude data of various filters $\mathbf{m_0}$ represent the redshift $z$. The redshift will be the result that maximises the likelihood. 

We chose PDFs with $100$ points and the same redshift range as the chosen for the other algorithms.

\subsection{ANNz2}

It is possible to obtain photo-zs through an Artificial Neural Network (ANN), a training-based method. It forms a map between the input variables and output ones, the connections through response functions are called neurons. \texttt{ANNz} \cite{collister2004annz} is a software that uses a multilayer perceptron (MLP).
Of the many layers of nodes, the inputs that are the magnitudes of $k$ filters, $\bf{m}_k$, are in the first layer, while the last ones contain the outputs: the photometric redshift $z_{p}$ and the PDFs.

To obtain $z_p$, the data must be divided into training, testing, and evaluation sets. For samples of galaxies without spectroscopic redshift, one needs to match the galaxies to a spectroscopic survey to use for training and testing.

What ANN does is minimise a cost function $E$ that compares the estimated $z_p$ according to the weight $\bf{w}$ and the input magnitudes, and the spectroscopic redshift $z_k$ of the training set. $E$ is written as follows:
\begin{equation}
    E = \sum_k  \left[z_{p}(\bf{w},\bf{m}_k)-z_k\right].
\end{equation}

\texttt{ANNz2} \cite{sadeh2016annz2} is an improved version to find the $z_p$ PDF. This can be done through a randomised regression that combines machine learning methods (MLM), which are trained and perform better. The MLM with the best performance is chosen as the final estimator. 

We used the Random Regression technique, dividing the reference set into four parts, the training with $20,608$ galaxies and three test sets each with $1,000$; $2,000$; and $2,760$ galaxies, respectively. An ensemble of $100$ random MLMs were used in the photo-z estimation, $90$ k-Nearest Neighbours (kNN), and $100$ PDF bins.

\subsection{DNF}

Another trained-based method is the nearest neighbours method. \cite{de2016dnf} developed two methods based on the k-nearest neighbour (kNN) algorithm. 
An Euclidean Neighbourhood Fitting (ENF) treats all the training galaxies as equally important, the estimated redshift will be close to where there are closer galaxies in the magnitude surface. A direction-oriented fitting, on the other hand, considers how the multi-magnitude surface looks depending on its direction, in this case, the direction is the estimated redshift. Imagine a single galaxy with many neighbours with their respective spectroscopic redshift. For a galaxy at the centre of its neighbours,  its redshift will point to the direction where the neighbours form a smoother surface. 

A good method ensures that the neighbours do not have just similar spectroscopic redshifts, but also a similar multi-magnitude surface. The Directional Neighbourhood Fitting (DNF) \cite{de2016dnf} provides a better redshift estimator compared to \texttt{ENF}.

The distance between two galaxies $D$ in a magnitude space is:
\begin{equation}
    D= \sqrt{\sum_i^n (m_i^t-m_i^p)^2},
\end{equation} where $m_i^t$ is the magnitude of the training sample and $m_i^p$ is the magnitude of the galaxies with redshift to be estimated, $n$ is the number of magnitude bands. This is the distance used with \texttt{ENF/DNF}.

For the DNF estimator, one might consider the angle $\alpha$ formed between two position vectors in magnitude space.
The perfect near-neighbour galaxy has $\alpha=0$, where the distance to a neighbour galaxy is $DN= D^2 \sin^2 \alpha$. We also chose 100 points to construct the PDFs. For the specifications with \texttt{ENF/DNF}, we chose 100 neighbours and ensured photometric redshifts remained inside the training redshift range.

\subsection{Spectroscopic sample}

\begin{figure}
    \centering
    \includegraphics[width=.8\linewidth]{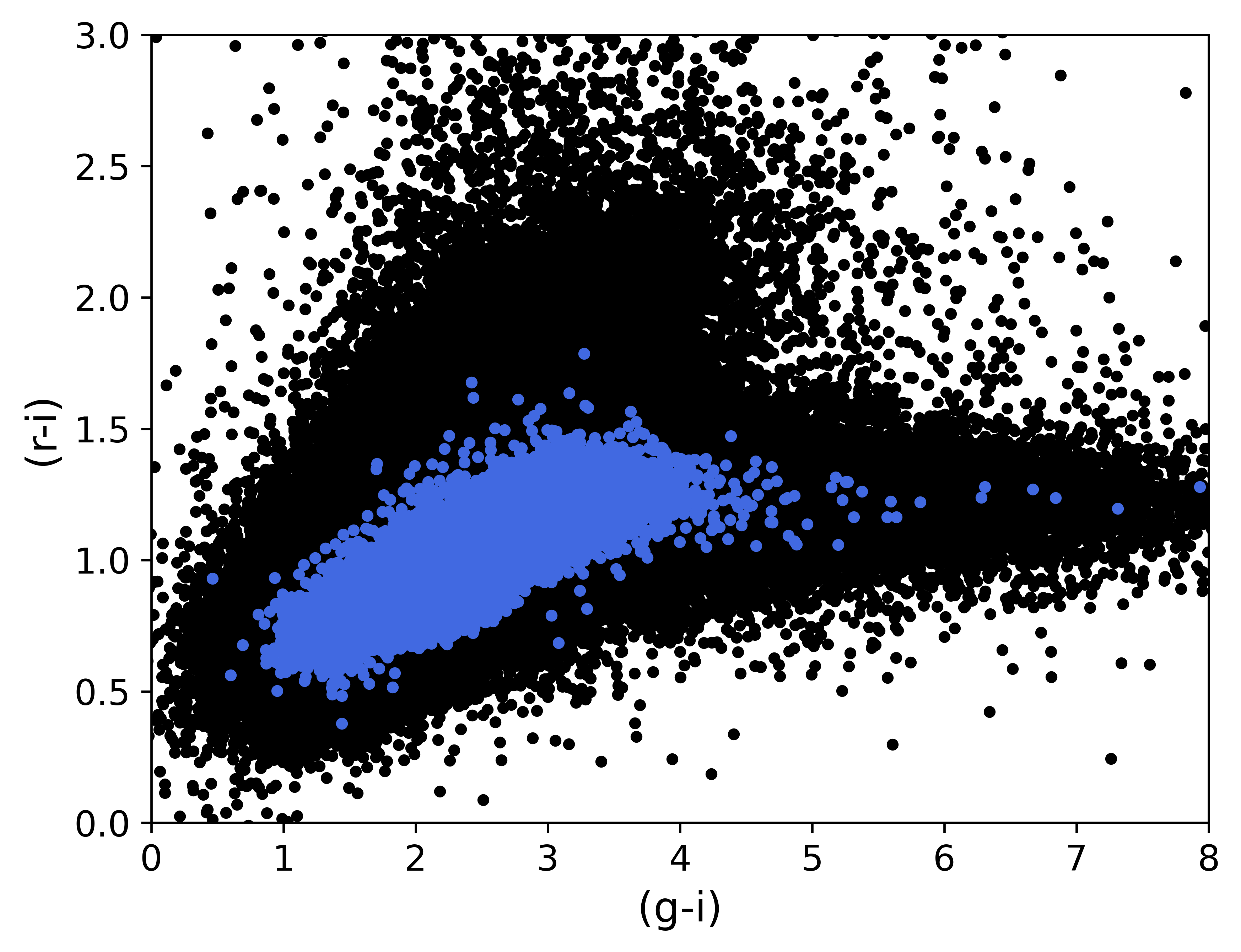}
    \caption{Colours $r-i$ vs $g-i$. of the DES galaxies (black) and the matched spec-z sample (blue).}
    \label{fig:filter-spec}
\end{figure}

\begin{figure}
    \centering
    \includegraphics[width=.8\linewidth]{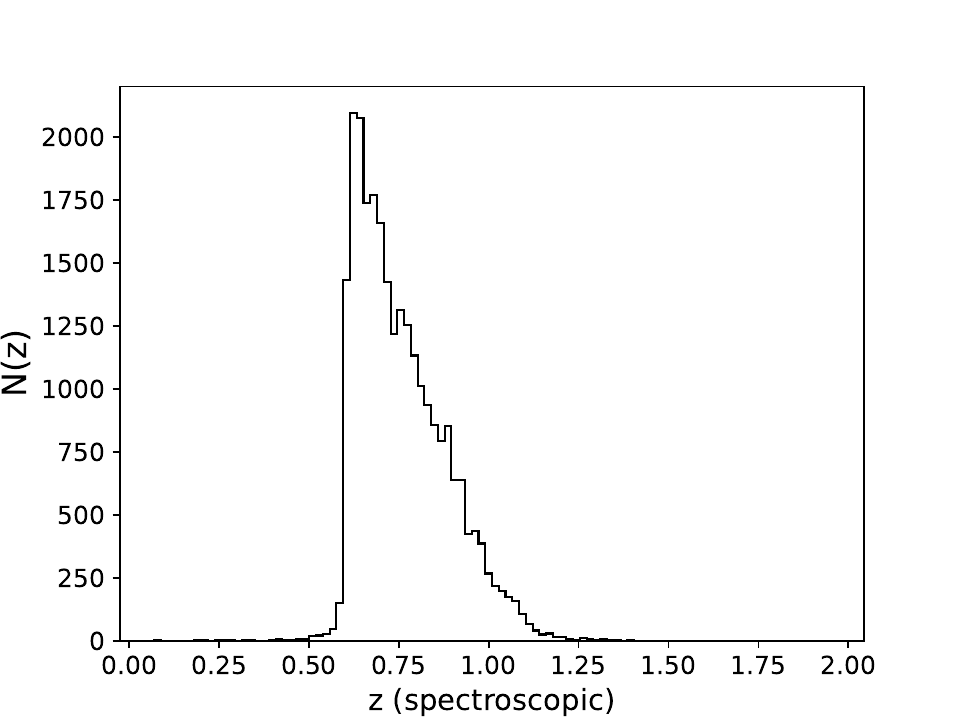}
    \caption{Spec-z sample distribution.}
    \label{fig:reference_sample}
\end{figure}

The reference spectroscopic galaxy set consists of $25,760$ galaxies from: the VIMOS Public Extragalactic Redshift Survey (VIPERS) Data Release 2 (PDR-2) (redshift flag Flag$\geq2$) \cite{scodeggio2018vimos} W1 and W4 equatorial fields, DEEP2 Galaxy Redshift Survey (Flag Q$>$3) \cite{newman2013deep2}, VIMOS VLT Deep Survey (VVDS) \cite{le2005vimos}, and the Sloan Digital Sky Survey (SDSS) eBOSS LRG pCMASS \cite{wang2020clustering} (HIZ\_LRG flag \cite{alam2015eleventh}) . 
These galaxies were found by matching their celestial coordinates with a tolerance of $2$ arcseconds.
We used the Random Regression technique, dividing the reference set into four parts, the training with $20,608$ galaxies and three test sets each with $1,000$; $2,000$; and $2,760$ galaxies, respectively. An ensemble of $100$ random Machine Learning Methods was used in the photo-z estimation, $90$ kNN, and $100$ PDF bins. In Figure~\ref{fig:filter-spec}, we show the distribution of colours $g-i$ and $r-i$ for the DES observations for its whole BAO sample in black and our matched galaxies in blue, we do not have information for $r-i>1.5$, but this is shown to be irrelevant in the results. Figure~\ref{fig:reference_sample} shows the redshift distribution of this sample, ideal for an LSS survey with LRG. 


 \texttt{BPZ} does not need a training set of spectroscopic redshift, thus the required input information was simply the magnitudes of the galaxies as well as the filters information, we added the DECam filters transmission as a function of wavelength to the \texttt{BPZ} inputs. We chose PDFs with $100$ points and the same redshift range as the chosen for the other algorithms.

We used the same spectroscopic sample described above for \texttt{DNF}/\texttt{ENF}. We also chose 100 points to construct the PDFs. For the specifications with \texttt{DNF}/\texttt{ENF}, we chose 100 neighbours and ensured photometric redshifts remained inside the training redshift range. Because \texttt{DNF}/\texttt{ENF} estimates the redshift according to neighbouring galaxies from the spectroscopic sample, we had to divide the spectroscopic set into $\sim 5000$ training galaxies and the remaining to compute the metrics for these two estimators, otherwise the result would be exactly the spec-z result. 

In Table~(\ref{tab:zeff}), we show the effective redshift $z_{\rm eff}$ defined in \cite{ross2017optimized} and the 68th percentile of the redshift error $\sigma_z$, defined as
\begin{equation}\label{eq:sigmaz}
    \sigma_z = \frac{(z_p-z)}{(1+z)}.
\end{equation}
The comparison was made with the spectroscopic dataset used from training and testing. For \texttt{ANNz2}, we used the set separated to test for the algorithm pipeline. The \texttt{BPZ} bias was calculated with all the available spectroscopic observations. Lastly, for \texttt{DNF}/\texttt{ENF}, we had to run the estimator with the galaxies of the training set, since \texttt{DNF}/\texttt{ENF} does not have a testing step. We separated 80\% for training and 20\% for testing \texttt{DNF}/\texttt{ENF}, and the 20\% was used to compute the bias between redshifts. \texttt{DNF}/\texttt{ENF}'s $\sigma_z^{68}$ matches the results obtained by \cite{san2024dark} in the incomplete training set case, which means we are providing sufficient spectroscopic reference galaxies. 

\begin{table}[ht]
    \centering
    \begin{tabular}{c|c|c}

       Data & $z_{\rm eff}$ & $\sigma_z^{68}$\\
       \hline
        \texttt{ANNz2} & 0.856 &$ ^{+0.007}_{-0.021}$\\ 
        \hline
        \texttt{BPZ} & 0.848 &$^{+0.009}_{-0.039}$\\
        \hline
        \texttt{ENF} & 0.854 &$^{+0.018}_{-0.016}$\\
        \hline
        \texttt{DNF} & 0.819 & $^{+0.054}_{-0.016}$\\

    \end{tabular}
    \caption{Uncertainty ($\sigma_z^{68}$) and effective redshift ($z_{\rm eff}$) of each algorithm using the full survey.}
    \label{tab:zeff}
\end{table}

\begin{figure}
    \centering
    \includegraphics[width=\linewidth]{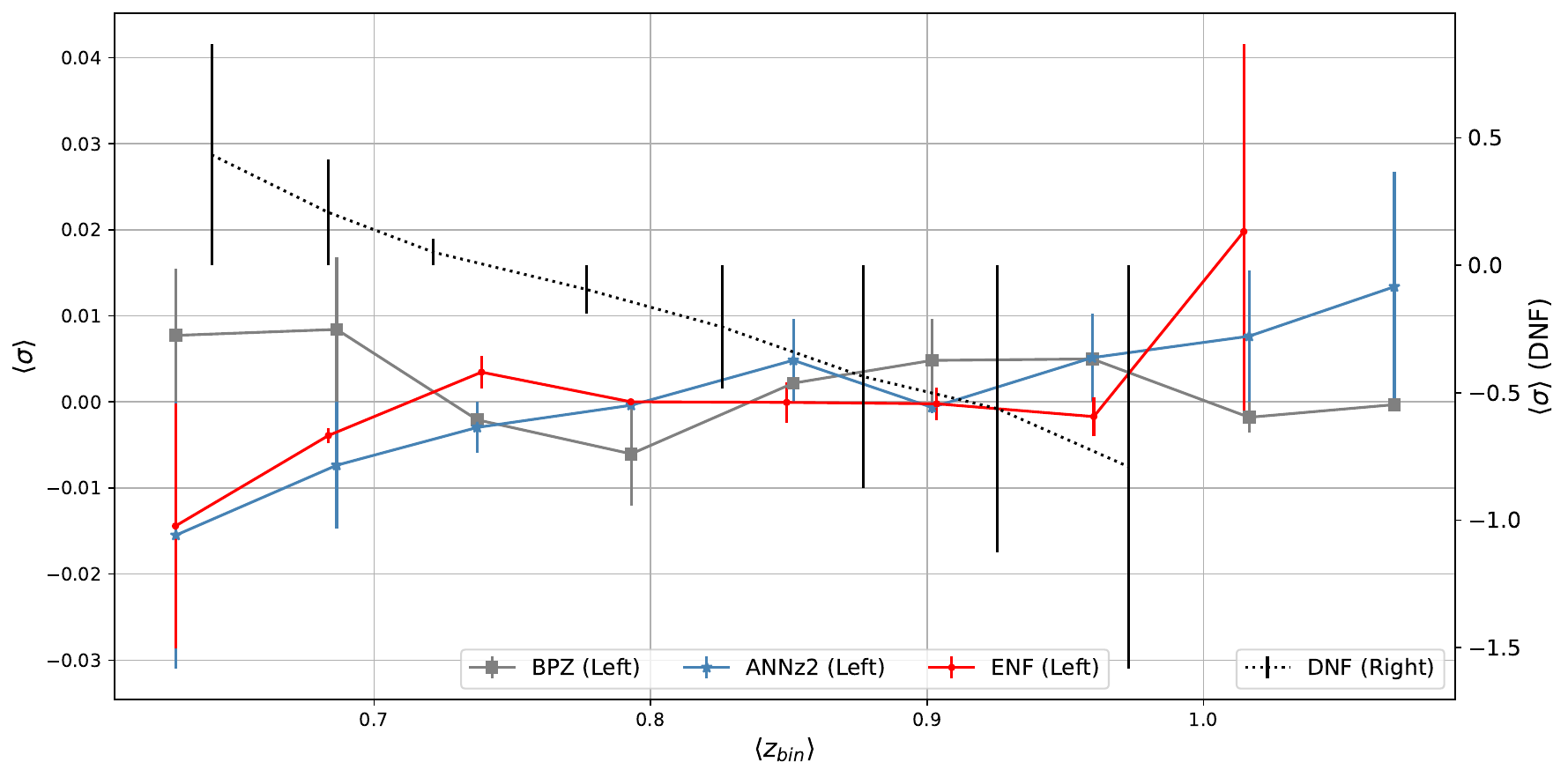}
    \caption{Redshift bias $\langle \sigma \rangle$ w.r.t. averaged redshift $\langle z_{bin}\rangle $ in different bins of each photo-z estimator. \texttt{ANNz2} (blue line with star symbols), \texttt{ENF}(red line with dots), \texttt{BPZ} (gray line with swuares), and \texttt{DNF} (dotted line). The \texttt{DNF}'s bias is shown in the right y-axis. The error bars were estimated from the 68th percentile of $\langle \sigma \rangle$ distribution around zero.}
    \label{fig:biases}
\end{figure}

\section{Bias per redshift bin}\label{sec:bias_bin}
A good way to compare these estimates is the quantity $\sigma_{68}$, which is the $68\%$ confidence region of the bias $\sigma$:
\begin{align}\label{eq:bias}
    \sigma =  z_{photo}-z_{train}.
\end{align}

We plot the bias, Eq.~(\ref{eq:bias}), per redshift bin in Figure~(\ref{fig:biases}) and their respective uncertainty based on the percentile 68 pra the bias distribution, the sample used to compare is the same as the previous plots. The error bars represent well the number of spectroscopic redshifts available per bin; they are smaller for bins with more spec-z. Consequently, the bias and errors are larger for higher redshift, a common lack of available data due to the nature of spectroscopic surveys. The red squares represent the \texttt{ENF} bias per bin, they are larger for $z>0.95$, this is expected once we do not have enough galaxies with these redshift values. The same is true for \texttt{ANNz2}, but the bias is smaller, indicating a better performance of this estimator. {\texttt{DNF} is the most biased estimator for most bins. Compared to the DES Collaboration results in \cite{sevilla2021dark}, we agree that in the bins around $z_p=0.8$, the bias increases at $z_p>1.0$. 

In Table~(\ref{tab:zeff}), we show the effective redshift $z_{\rm eff}$ described by \cite{ross2017optimized}, and the 68th percentile of the redshift error $\sigma_z$ \ref{eq:sigmaz}. This first result is based on the full sample.
The comparison was made with the spectroscopic data set used in the training and testing. For \texttt{ANNz2}, we used the separated set to test the algorithm pipeline. The \texttt{BPZ} bias was calculated with the available spectroscopic observations. Lastly, for \texttt{ENF/DNF}, we had to run the estimator with the galaxies of the training set, since \texttt{ENF/DNF} does not have a testing step. We separated 80\% for training and 20\% for testing \texttt{ENF/DNF}, the 20\% was used to compute the bias between redshifts.

\section{Results}\label{sec:results}

Our results for the four methods are available in the Zenodo repository with DOI: \href{10.5281/zenodo.14290701}{10.5281/zenodo.14290701}. In Figure~\ref{fig:anz2disp}, we show the \texttt{ANNz2} testing sample compared to the resulting photo-z. The cyan dotted line shows the $\sigma_{99}$ error, the dashed blue line represents the $\sigma_{68}$ error with the expected result. The black and white pixels represent the number of galaxies concentrated with that particular spec-z/photo-z value. The resulting redshift estimation was satisfactory, the sample distribution matches the spectroscopic redshift pattern without much dispersion. 

\begin{figure}
    \centering
    \includegraphics[width=.95\textwidth]{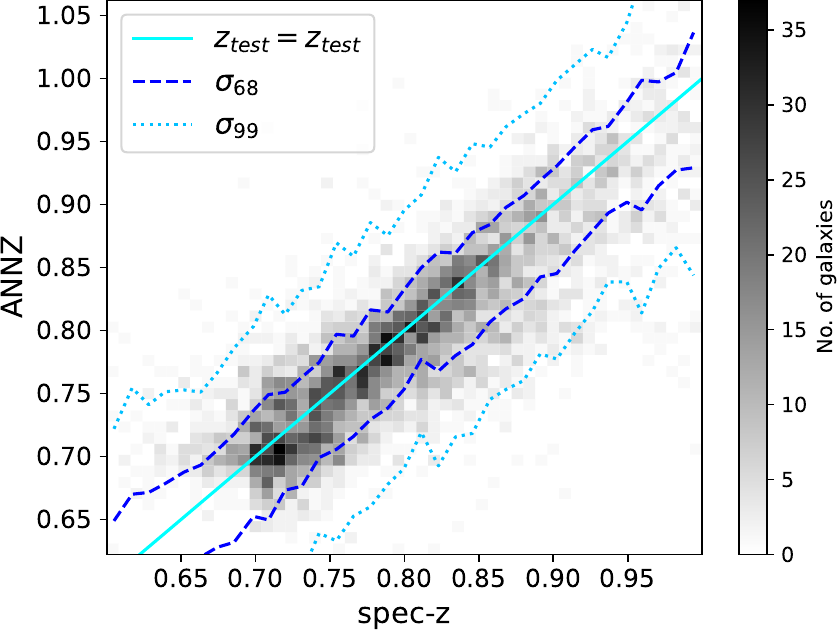}
    \caption{\texttt{ANNz2} photo-z compared to the spectroscopic training set. The horizontal axis is the spec-z testing sample. The colorbar represents the number of galaxies in a 2D histogram. The cyan solid-line is the expected linear relation between photo-z and spec-z. The blue dashed-line represents the 68\% confidence level in a $\sigma_z^{68}$ distribution, and the cyan-dotted one the 99\% confidence level.}
    \label{fig:anz2disp}
\end{figure}

Figure~\ref{fig:bpz_vipers} shows the high bias between the \texttt{BPZ} photo-z and the matched samples $spec-z$. Compared to \texttt{ANNz2}, this estimator has a bigger error, indicated by the wider $\sigma_{68}$ region. The results are consistent with the expected redshift, with little dispersion from the $z_p=z_{spec}$ line. For redshifts greater than 1.1, $\sigma_{99}$ deviates from a straight line. The stripped pattern is due to the limitation of the estimator of setting a redshift resolution, we chose a resolution of $0.02$. This same pattern can be seen in other studies using \texttt{BPZ}, like in \cite{margoniner2008photometric}.
 \begin{figure}
     \centering     \includegraphics[width=1\textwidth]{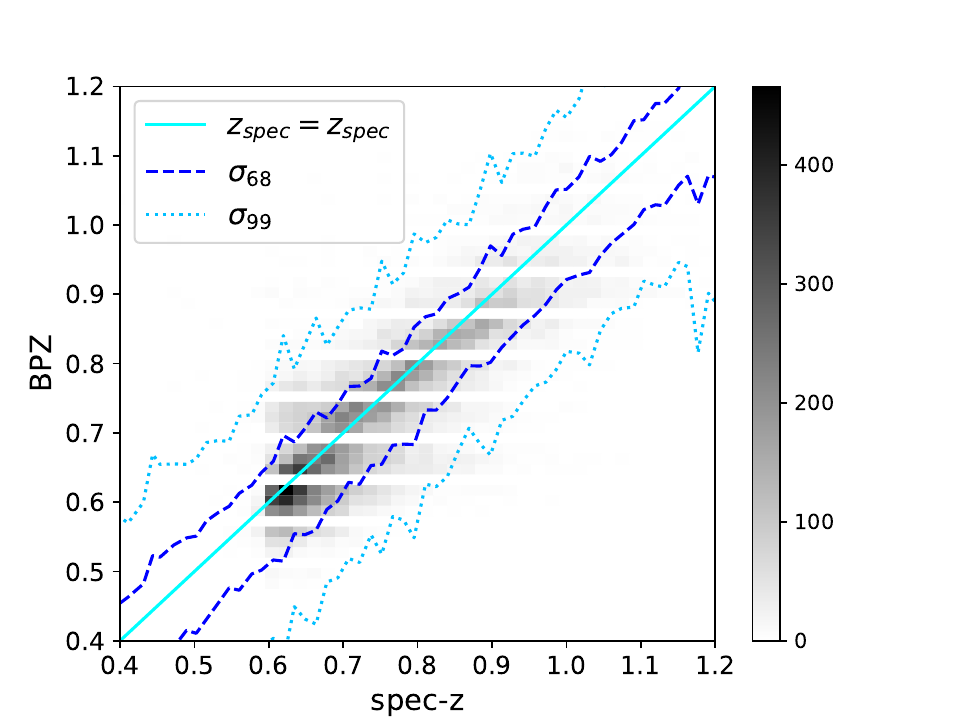}
     \caption{\texttt{BPZ} vs spec-z full reference set. The colorbar represents the number of galaxies in a 2D histogram. The cyan solid-line is the expected linear relation between photo-z and spec-z. The blue dashed-line represents the 68\% confidence level in a $\sigma_z^{68}$ distribution, and the cyan-dotted one the 99\% confidence level.}
     \label{fig:bpz_vipers}
 \end{figure}

 \begin{table}
    \centering
    \begin{tabular}{|c|c|c|c|c|c}
        \hline
        Paper & Algorithm & $\sigma_{68}$ & Sample &  Training set size\\
        \hline
        DES SV \cite{sanchez2014photometric} & \texttt{ANNz2} (test 1 cut) & $\sim$0.049 & DES SV  & 7,249\\
        DES SV \cite{sanchez2014photometric} & \texttt{BPZ} (test 1) & $\sim$0.056 & DES SV  & - \\
        DES SV \cite{sanchez2014photometric} & \texttt{DESDM} (test 1) & $\sim$0.055 & DES SV  & 7,249 \\
        DES Deep Field \cite{san2024dark} & \texttt{DNF} Complete & 0.0373 & DESY3 Deep Field & 26,883\\
        DES Deep Field \cite{san2024dark} & \texttt{DNF} Incomplete &  0.0500 & DESY3 Deep Field & 22,651\\
        \hline
    \end{tabular}
    \caption{Photo-z estimation using DES objects.}
    \label{tab:refs}
\end{table}

 The \texttt{ENF} results are shown in Figure~\ref{fig:enf2disp}. This time the deviation in the errors region happens at $z\sim0.95$, this indicates that the estimator is not accurate enough for higher redshift values. Finally in Figure~\ref{fig:dnf2disp}, \texttt{DNF} seems to overestimate the redshift and has the same issue at higher redshift as \texttt{ENF}. Both algorithms show an asymmetric distribution, the negative $\sigma_{68}$ is closer to the $z_{spec}=z_{p}$ function for the $z<0.9$, these algorithms are overestimating the galaxy's photo-z. A similar discrepancy appeared in the lower redshift region in \cite{reis2012sloan} for the Sloan Digital Sky Survey photometric catalog. 
 
 In table~\ref{tab:refs}, we included results using DES data with different algorithms, from the Scientific Verification and Deep Field. We had to estimate the $\sigma_{68}$ from \cite{sanchez2014photometric} to the equivalent version of our definition \footnote{Considering their mean spec-z is around 0.75, we divide their $\sigma_{68}$ by (1+z).}. Our results show lower $\sigma_{68}$ than most cases. The exception is \texttt{DNF} Complete training (all the available filters from DECAM) set from \cite{san2024dark}. Compared to \cite{san2024dark}'s \texttt{DNF} incomplete case (only $g,r,i,z$ filters like the BAO sample), we found compatible results.

\begin{figure}
    \centering    \includegraphics[width=1\textwidth]{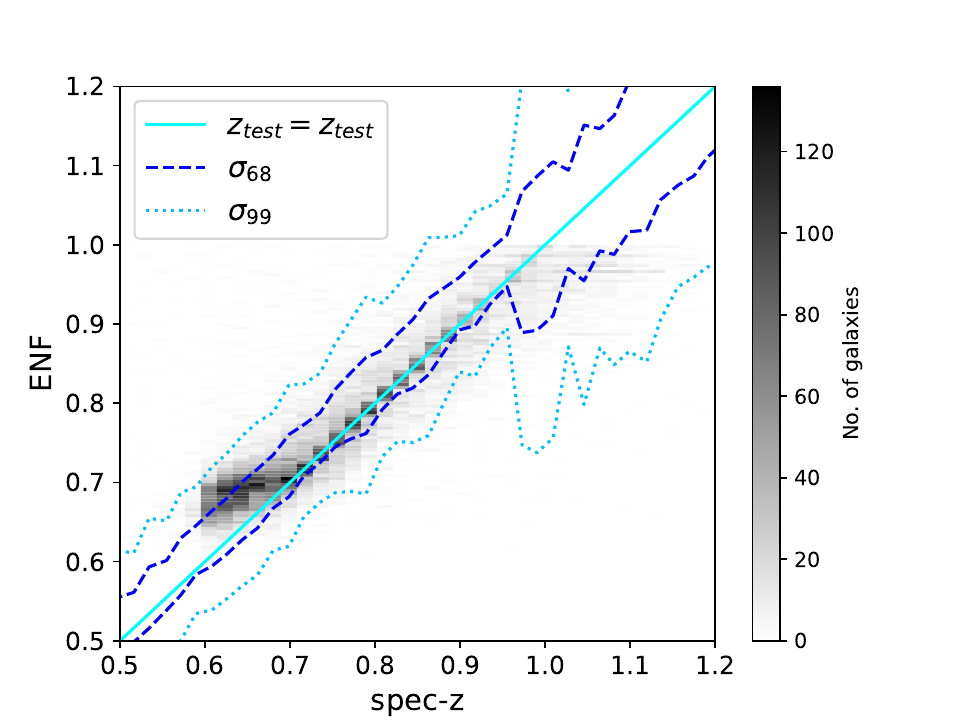}
    \caption{\texttt{ENF} photo-z vs spec-z. The colorbar represents the number of galaxies in a 2D histogram. The cyan solid-line is the expected linear relation between photo-z and spec-z. The blue dashed-line represents the 68\% confidence level in a $\sigma_z^{68}$ distribution, and the cyan-dotted one the 99\% confidence level.}
    \label{fig:enf2disp}
\end{figure}

\begin{figure}
    \centering    \includegraphics[width=1\textwidth]{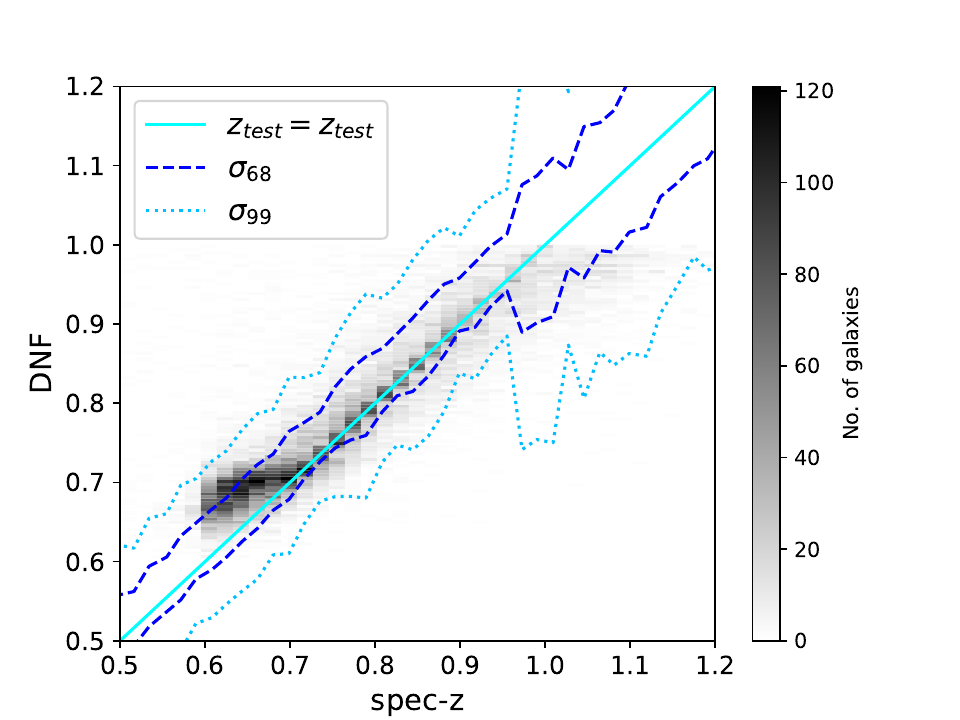}
    \caption{\texttt{DNF} photo-z vs spec-z.  The colorbar represents the number of galaxies in a 2D histogram. The cyan solid-line is the expected linear relation between photo-z and spec-z. The blue dashed-line represents the 68\% confidence level in a $\sigma_z^{68}$ distribution, and the cyan-dotted one the 99\% confidence level.}
    \label{fig:dnf2disp}
\end{figure}

In order to make sure our spectroscopic sample was enough for \texttt{DNF}, we compare, in Figure \ref{fig:dnf_compared}, our result in the x-axis with the BAO sample by the DES Y3 \cite{abbott2022dark} in the y-axis. They are highly correlated for all the redshift values. The DES Collaboration used $\sim 2.2\times10^5$ spectroscopic matched galaxies in their estimation
spectra matched \cite{abbott2022dark}. Despite their superior access to spectroscopic data, our analysis is comparable with $\langle \sigma \rangle =\langle z_{Pz \,\,cats}-z_{DESY3} \rangle= - 0.003$. Figure \ref{fig:dnf_compared} shows the 68th percentile for this bias with the same colour scheme used above. The results diverge for $z>0.9$, this means \texttt{DNF} is more sensitive to the training set in this range.

\begin{figure}
    \centering
    \includegraphics[width=.8\linewidth]{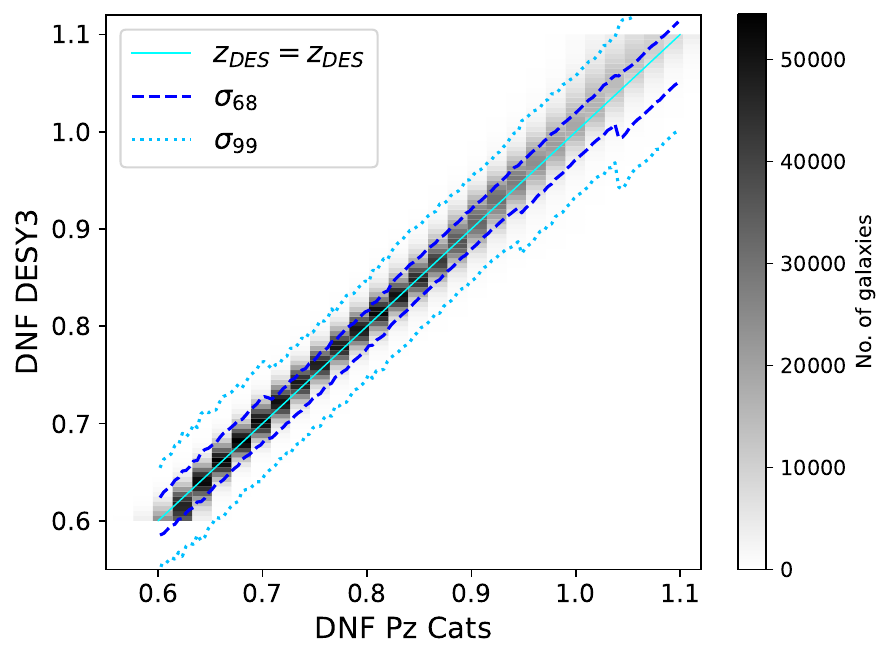}
    \caption{\texttt{DNF} results with our training set vs the DESY3 one. The colorbar represents the number of galaxies in a 2D histogram. The cyan solid-line is the expected linear relation between the two data sets. The blue dashed-line represents the 68\% confidence level in a $z_{DES} - z_{PzCats }$ distribution, and the cyan-dotted one the 99\% confidence level.}
    \label{fig:dnf_compared}
\end{figure}

\subsection{Consistency tests based on the redshift Probability Distribution Function}

Due to the lack of a bigger number of filters, we may find redshift Probability Distribution Functions (PDF) that are not perfect, such as a Dirac delta function. An accurate but not free-of-systematics PDF ideally looks like a Gaussian distribution, and the realistic case results in a survey with nearly Gaussian and multi-modal PDFs. 

\begin{table}[ht]
    \centering
    \begin{tabular}{c|c|c|c|c|c|c}
       Estimator& Method  & No. of galaxies & $z_{\rm eff}$ &$\sigma_z^{68}$ & Area (deg$^2$) & Galaxies/deg$^2$\\
       \hline
        \texttt{ANNz2} &  Gaussian & $2,931,677$& $0.817$& $^{+0.014}_{-0.031}$&  21,065.36& 139.17\\
        \hline
        \texttt{ANNz2} & Small peaks & $5,133,775$ & $0.799$ &$^{+0.009}_{-0.022}$& 21,065.36& 243.71\\
        \hline
        \texttt{ENF} & Gaussian & $299,427$ & $0.834$ & $^{+0.040}_{-0.084}$ &  21,032.17 & 14.24\\
        \hline
        \texttt{ENF} & Small peaks & $2,256,616$ & $0.768$ & $^{+0.046}_{-0.055}$ & 21,065.36& 107.12\\
        \hline
        \texttt{DNF} & Gaussian & $2,111,461$ &0.868 & $^{+0.116}_{-0.121}$ &  21,078.79& 100.17\\
        \hline
        \texttt{DNF} & Small peaks &$2,456,190$ & $0.761$ & $^{+0.033}_{-0.142}$ &21,078.79 & 116.52\\
        \hline
        \texttt{BPZ} & Gaussian & $3,973,671$ & $0.812$ & $^{+0.016}_{-0.045}$ & 21,065.36& 188.64\\
        \hline
        \texttt{BPZ} & Small peaks & $5,904,860$ & $0.811$ & $^{+0.008}_{-0.042}$ & 21,065.36& 280.31
    \end{tabular}
    \caption{All the cuts for each estimator.}
    \label{tab:cuts}
\end{table}

Unfortunately, one cannot eliminate all the objects with secondary peaks, but it is possible to exclude the most noisy ones. Here, we computed the number of peaks per PDF and selected galaxies that contain a peak that cannot be larger than 30\% of the main peak. Furthermore, we selected PDFs close to a Gaussian distribution. We calculated the statistical moments $\mu_n$ using the distributions $PDF(z_p)$:
\begin{equation}
    \mu_n = \int\limits_{-\infty}^{\infty} z_p^n PDF(z_p) \dd z_p,
\end{equation}
where $n$ is the moment ordinal, we use the second moment to classify the distribution. For a Gaussian, the second moment is the sum of the average value squared and the variance ($\mu^2+\sigma^2$)\footnote{The code is available at: \url{https://github.com/psilvaf/bao_pz}.}. The mode and the mean are the same in a normal distribution, so we use the $z_p$ output of the estimators as $\mu$ and its respective error as $\sigma$. This sample we call Gaussian + "estimator". We shall call Small Peaks the case where the second biggest maximum is not bigger than thirty per cent of the global maximum, this guarantees that the noise is not dominant in the PDF. 

Contrasting with the reweighing approach outlined in \cite{lima2008estimating,cunha2009estimating}, which constructs the probability density function (PDF) of the sample by using the weights while also avoiding both the need for trimming and the evaluation of the PDFs for individual galaxies, our sample selection methodology is specifically structured to remove unreliable individual PDFs that might exist within the survey dataset. These PDFs are prone to noise due to their under-representation in the magnitude space, leading to potential distortions in data interpretation.

The resulting sub-samples are shown in the table \ref{tab:cuts}. \texttt{BPZ} has the biggest number of selected objects in both criteria. However, because of the Bayesian method, this estimator forces smoother PDFs, but that may not mean it is the best representation of all galaxies, we need to trust that the template available is representative enough of the galaxy's photometry. \texttt{ANNz2} shows better precision despite not having a sample as big as \texttt{BPZ}'s. \texttt{ENF} has the worst selection of Gaussian PDFs, with very few galaxies left for an LSS study. Lastly, the worst in precision is \texttt{DNF} but with sufficient galaxies for cosmological analysis. The Small Peaks results are more populated, this time \texttt{ANNz2} has better precision and more objects than the other algorithms. Choosing the least noisy PDFs may be an advantage to LSS studies for being less selective because the decrease in the number of objects is lower. 

The next question is to determine if the selected samples continue to represent robust large-scale structure (LSS) datasets. In Table~\ref{tab:cuts}, we have presented both the residual survey area and the density of objects, expressed as the number of galaxies per square degree. It is important to highlight that the spatial footprint of the survey has remained consistent, unaffected by the decrease in the quantity of objects. Nonetheless, there is a crucial caveat regarding the \texttt{ENF} Gaussian sample: the substantial reduction in object count for this particular sample renders it unsuitable for further analyses involving angular power spectra or two-point correlation functions. Our analysis of the galactic distribution through various cuts supports the data in Table~\ref{tab:cuts}. The \texttt{ENF} approach shows a decrease in galaxy counts, while an optimal sky direction features increased galaxy density linked to the spec-z sample. In contrast, \texttt{ANNz2} exhibits the smallest sky distribution post-cuts.

The principal aim of this survey is to identify the BAO signal. However, a reduction in the number of galaxies compromises the signal, and the of this reduction on statistical power in \cite{Ferreira:2025yvb}. It is important to highlight that BAO detection can be accomplished using two methodologies: one that operates in three dimensions and another that is transverse to the line of sight. To achieve accurate transverse BAO measurements, it is crucial to have an adequate density of galaxies per square degree. In contrast, for three-dimensional observations, the redshift distribution needs to be robust enough to capture the characteristic feature along the line of sight, irrespective of whether the survey footprint encompasses an extensively filled section of the sky. In particular, the most recent findings from the Dark Energy Spectroscopic Instrument (DESI) \cite{karim2025desi} provide an exemplary estimation of 3D BAO even when significant gaps are present across all its tracers. In the last column of Table~\ref{tab:cuts}, we included the number of galaxies per deg$^2$, the only sample to lose statistics considerably is \texttt{ENF}, all the others are capable of being a BAO sample.

\begin{figure}
    \centering
    \begin{subfigure}[b]{0.49\linewidth}
        \includegraphics[width=\linewidth]{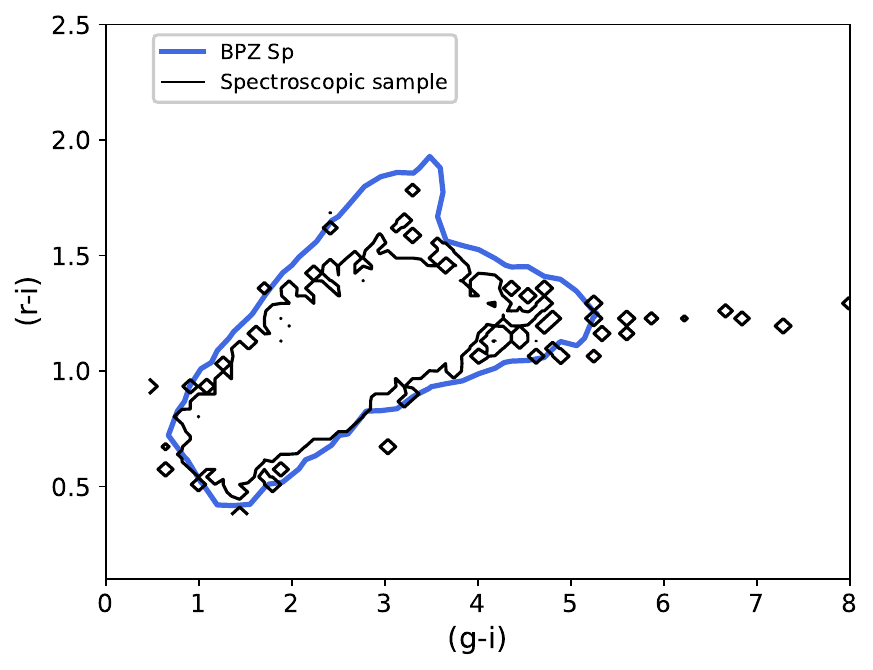}
        \caption{BPZ}
        \label{fig:bpz}
    \end{subfigure}
    \hfill
    \begin{subfigure}[b]{0.49\linewidth}
        \includegraphics[width=\linewidth]{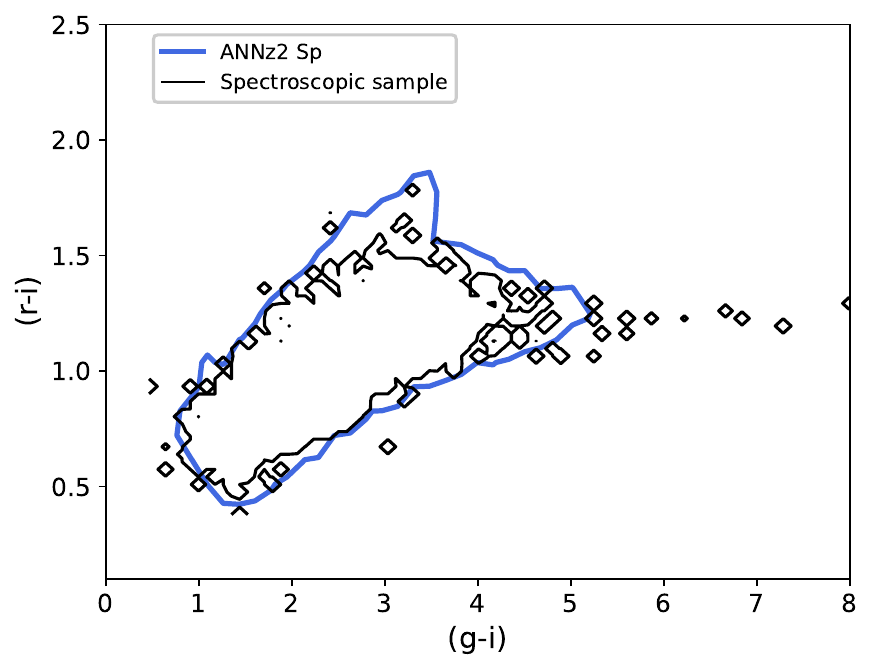}
        \caption{ANNz2}
        \label{fig:annz}
    \end{subfigure}
    
    \vspace{0.5cm} 
    
    \begin{subfigure}[b]{0.49\linewidth}
        \includegraphics[width=\linewidth]{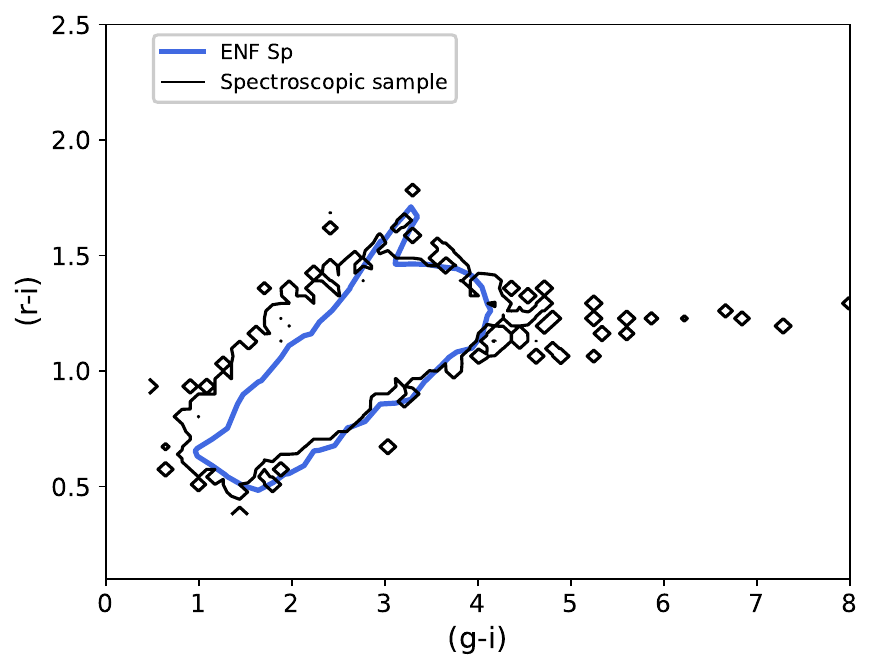}
        \caption{ENF}
        \label{fig:enf}
    \end{subfigure}
    \hfill
    \begin{subfigure}[b]{0.49\linewidth}
        \includegraphics[width=\linewidth]{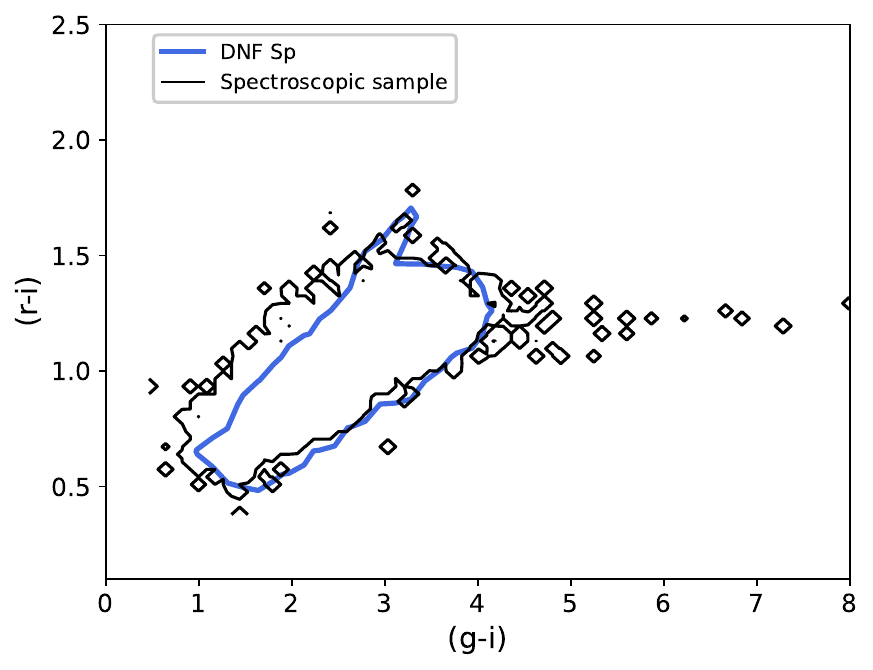}
        \caption{DNF}
        \label{fig:dnf}
    \end{subfigure}
    
    \caption{Comparison of colour distributions for different methods of photo-z (black) estimation and the matched spec-z sample (blue). The contours represent the $68\%$ confidence level of the distribution: (a) BPZ, (b) ANNz2, (c) ENF, and (d) DNF.}
    \label{fig:color_panel}
\end{figure}

\begin{figure}
    \centering
    \begin{subfigure}[b]{0.49\linewidth}
        \includegraphics[width=\linewidth]{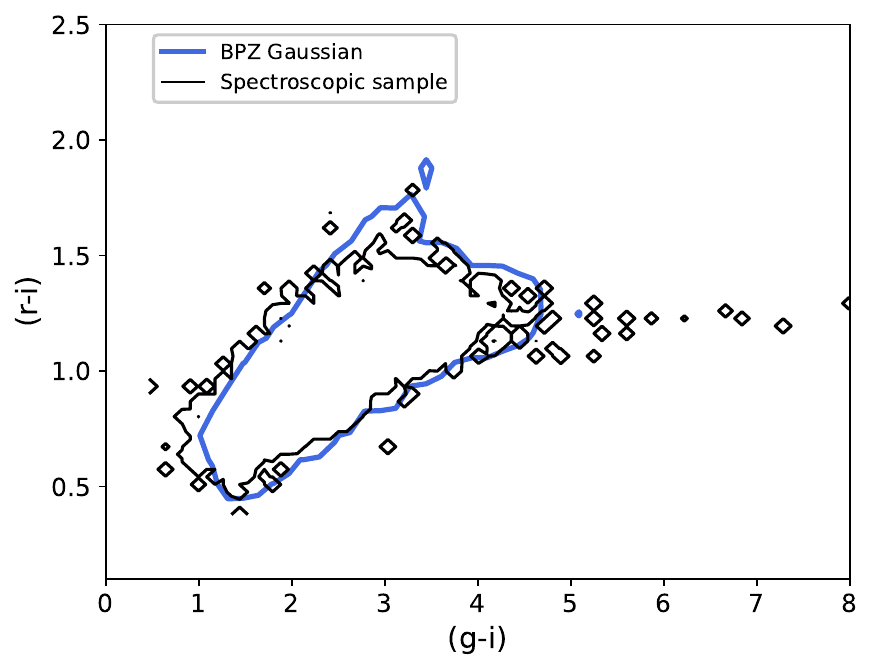}
        \caption{BPZ}
        \label{fig:bpz2}
    \end{subfigure}
    \hfill
    \begin{subfigure}[b]{0.49\linewidth}
        \includegraphics[width=\linewidth]{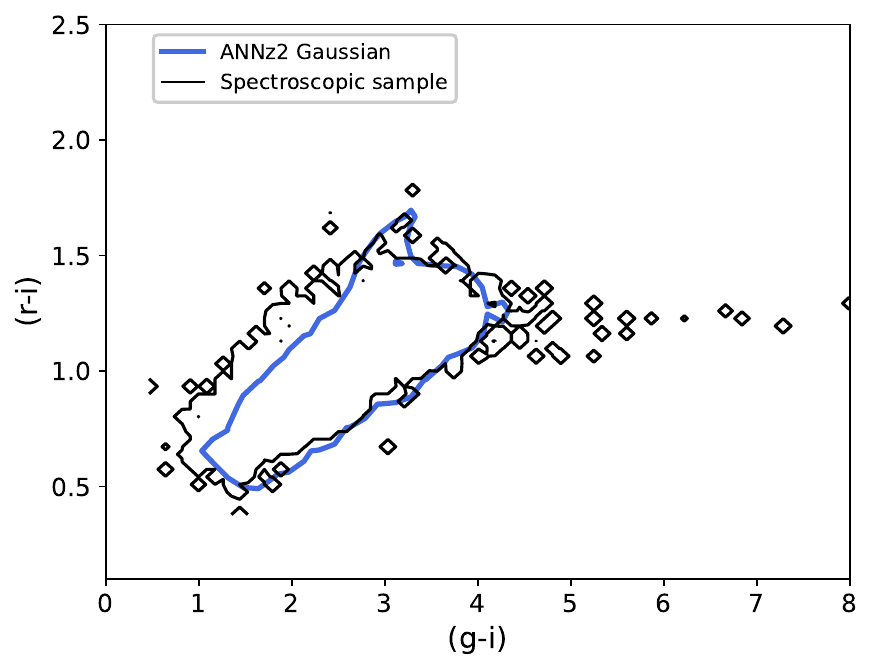}
        \caption{ANNz2}
        \label{fig:annz2}
    \end{subfigure}
    
    \vspace{0.5cm} 
    
    \begin{subfigure}[b]{0.49\linewidth}
        \includegraphics[width=\linewidth]{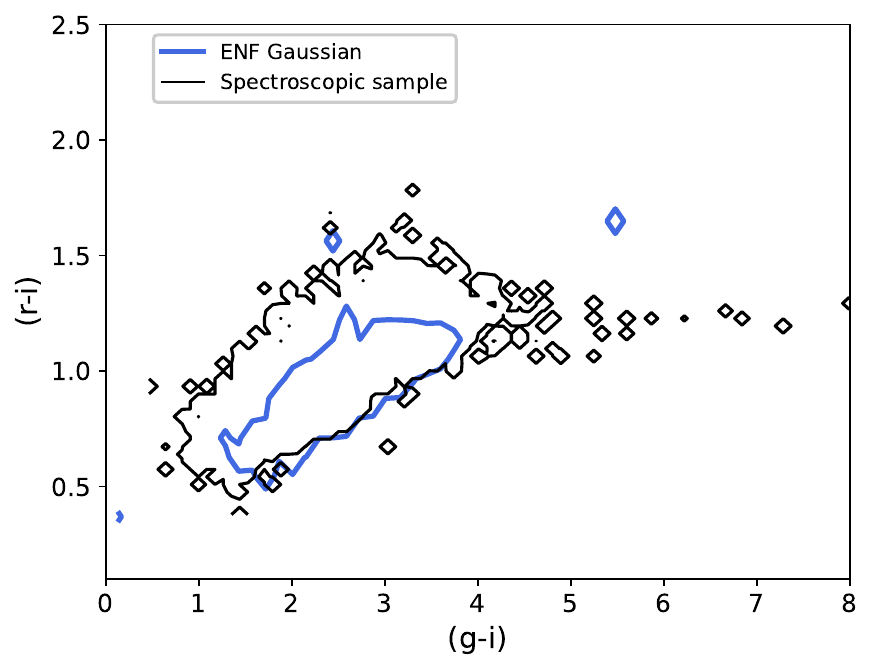}
        \caption{ENF}
        \label{fig:enf2}
    \end{subfigure}
    \hfill
    \begin{subfigure}[b]{0.49\linewidth}
        \includegraphics[width=\linewidth]{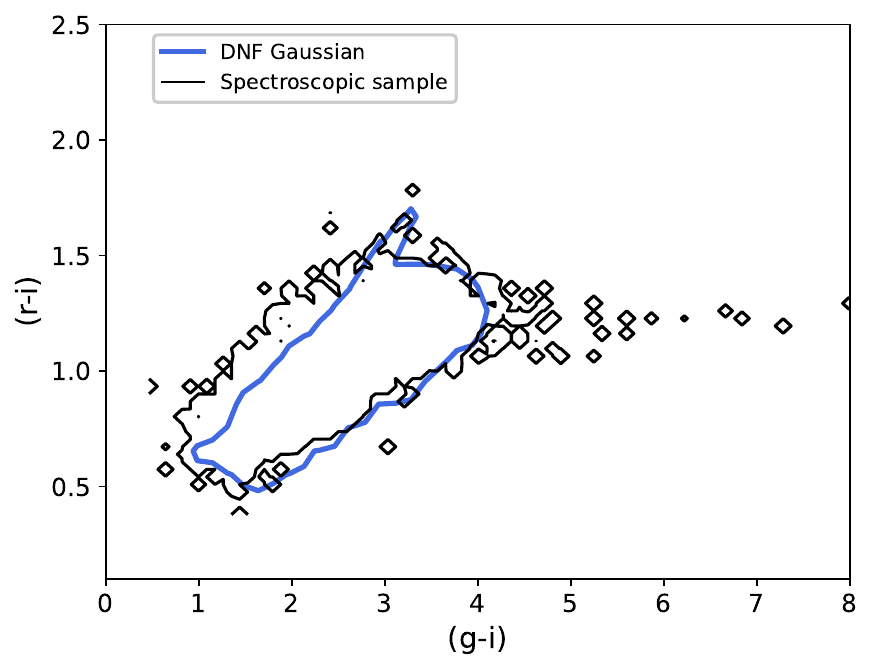}
        \caption{DNF}
        \label{fig:dnf2}
    \end{subfigure}
    
    \caption{Comparison of colour distributions for different methods of photo-z (black) estimation and the matched spec-z sample (blue). The contours represent the $68\%$ confidence level of the distribution: (a) BPZ, (b) ANNz2, (c) ENF, and (d) DNF. }
    \label{fig:color_panel2}
\end{figure}

\begin{figure}
    \centering
    \includegraphics[width=\linewidth]{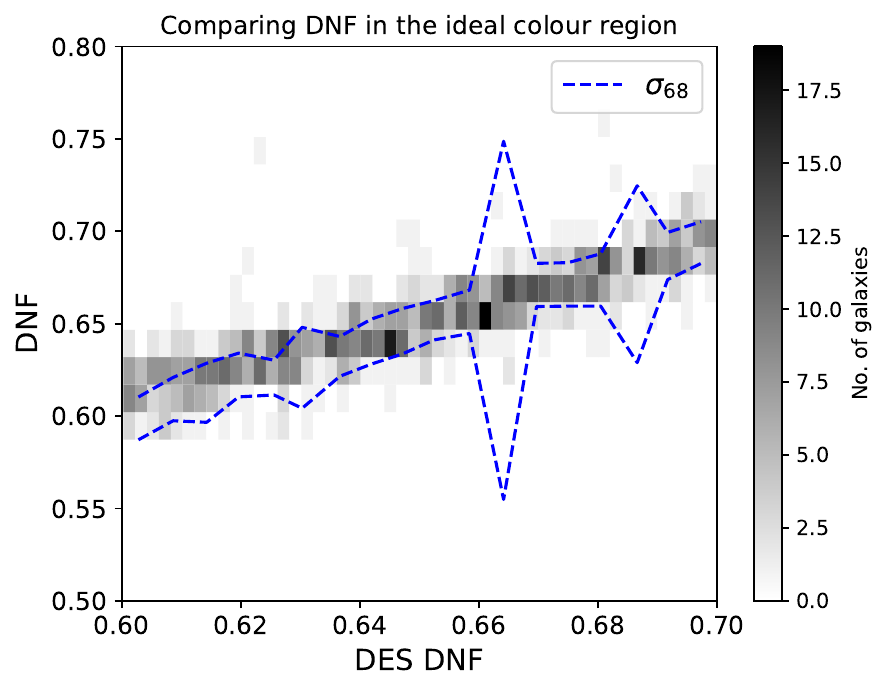}
    \caption{\texttt{DNF} results with our training set vs the DESY3 one inside the 1$\sigma$ color distribution of the spectroscopic sample. The colorbar represents the number of galaxies in a 2D histogram. The blue dashed-line represents the 68\% confidence level.}
    \label{fig:dnf_des_colors}
\end{figure}

From eq.~(\ref{eq:bias}) 68th percentile, we plot the distribution of $\sigma_z$ for all estimators and sample cuts. In Figure~\ref{fig:comparison_sigma}, we show the distribution of $\sigma_z$ for all the samples described here. The black line represents the whole survey, the blue one represents the samples only with Gaussian PDFs, and the grey area includes multimodal PDFs but not too noisy. 

\begin{figure}[ht]
    \centering
    \begin{subfigure}[b]{0.38\textwidth}
        \centering
        \includegraphics[width=\linewidth]{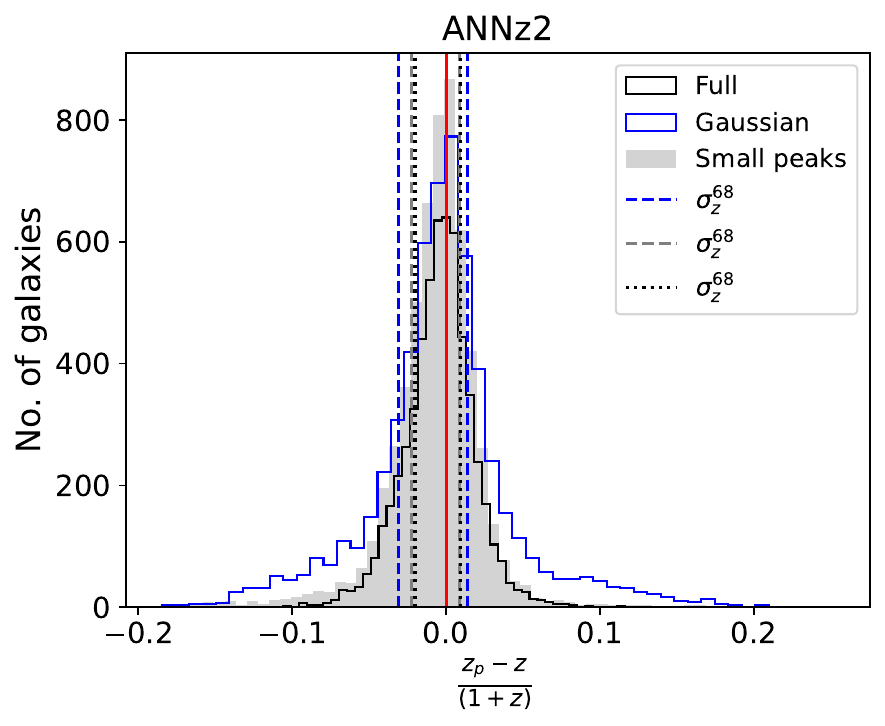}
        \caption{ANNZ $\sigma_z$}
        \label{fig:annz_hist_sigma}
    \end{subfigure}
    \begin{subfigure}[b]{0.4\textwidth}
        \centering
        \includegraphics[width=\linewidth]{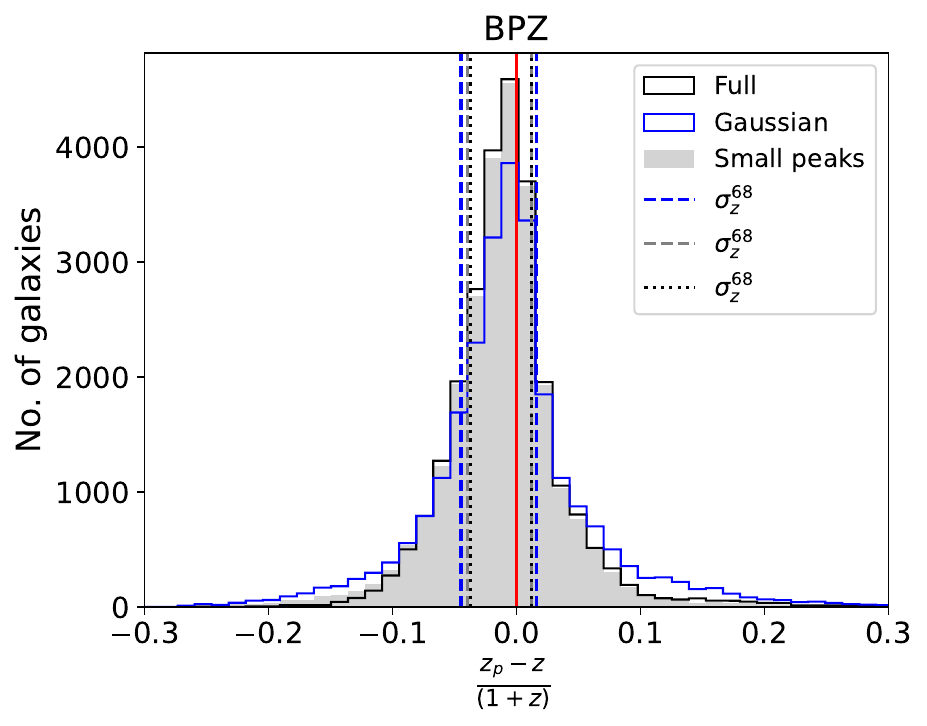}
        \caption{BPZ $\sigma_z$}
        \label{fig:bpz_hist_sigma}
    \end{subfigure}
    \hfill
    \begin{subfigure}[b]{0.38\textwidth}
        \centering
        \includegraphics[width=\linewidth]{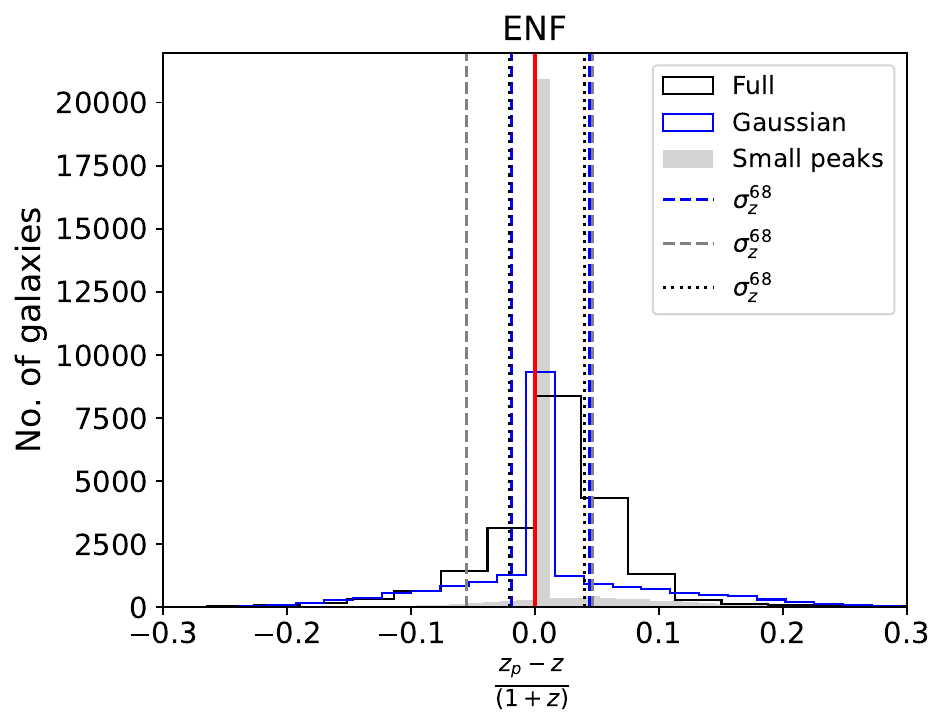}
        \caption{ENF $\sigma_z$}
        \label{fig:enf_hist_sigma}
    \end{subfigure}    
    \begin{subfigure}[b]{0.4\textwidth}
        \centering
        \includegraphics[width=\linewidth]{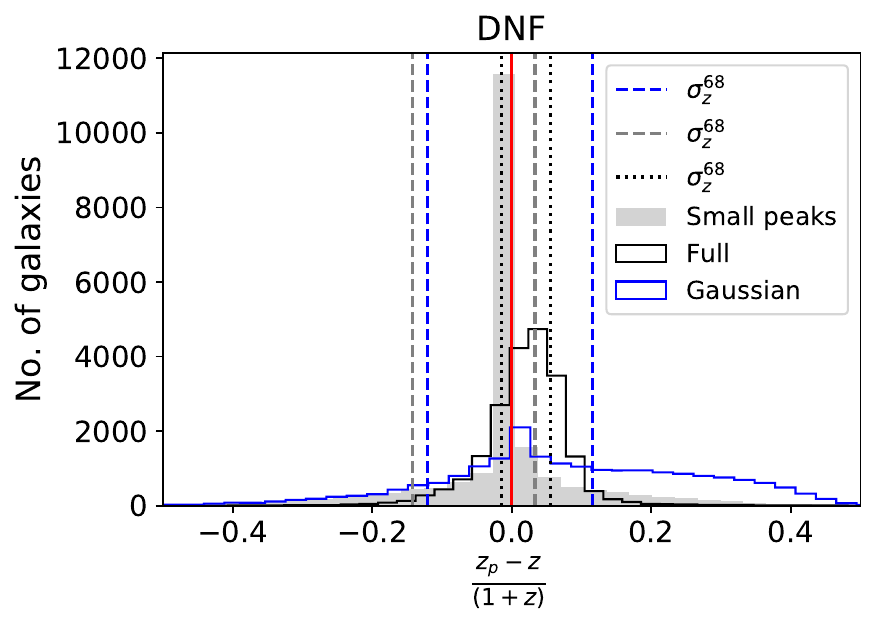}
        \caption{DNF $\sigma_z$}
        \label{fig:dnf_hist_sigma}
    \end{subfigure}    
    \caption{Comparison of $\sigma_z$ values from different models/algorithms. The vertical red-line marks the null difference, the black histogram has all galaxies, gray histogram represents the Small peaks cuts, and the blue histogram the Gaussian one with the corresponding colours of dashed vertical lines representing the 68\% confidence level.}
    \label{fig:comparison_sigma}
\end{figure}

The first panel has the \texttt{ANNz2} normalised bias, Figure~\ref{fig:annz_hist_sigma}, the vertical lines respecting the distribution colours. The expected normalised bias is a Gaussian distribution with mean zero, the three cases show the centres close to zero, but the distributions are not perfectly Gaussian, and the $\pm \sigma_{68}$ are not symmetric. The least accurate sample is the blue (Gaussian) one, the main reason is the loss of statistics. In Figure~\ref{fig:bpz_hist_sigma}, \texttt{BPZ}'s three distributions are very similar and not Gaussian. \texttt{ENF} and \texttt{DNF} (Figure~\ref{fig:enf_hist_sigma},\ref{fig:dnf_hist_sigma}) have the worst normalised bias distribution and biggest $\pm \sigma_{68}$.

\subsection{Precision with respect to the redshift}

Complementary to the results above, we show $\sigma_{68}$ w.r.t. the spectroscopic redshift of the matched galaxies used for testing (all galaxies for \texttt{BPZ}). The result is shown in Figure~\ref{fig:sigma_spec_z}. \texttt{BPZ} in figure~\ref{fig:bpz-sigma68} looses precision when $z>1.0$, for \texttt{ENF/DNF}(figs.~\ref{fig:enf-sigma68},\ref{fig:dnf-sigma68}) this happens at $z>0.9$. The only photo-z estimator to show consistent and precise results is \texttt{ANNz2}. 

\begin{figure}
    \centering
    \begin{subfigure}[b]{0.45\linewidth}
        \centering
        \includegraphics[width=\linewidth]{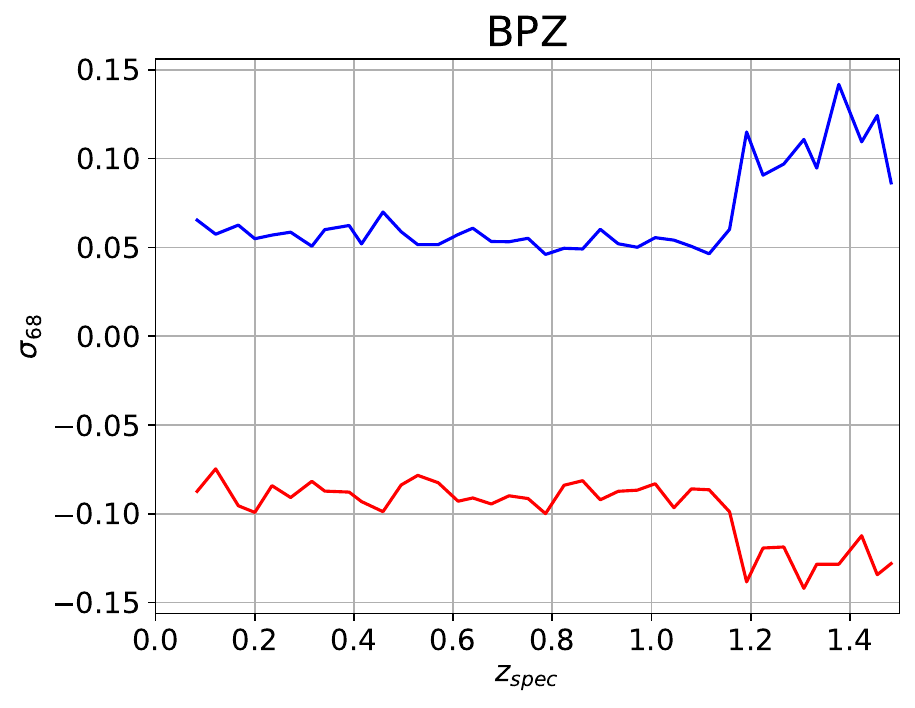}
        \caption{\texttt{BPZ} }
        \label{fig:bpz-sigma68}
    \end{subfigure}
    \hfill
    \begin{subfigure}[b]{0.45\linewidth}
        \centering
        \includegraphics[width=\linewidth]{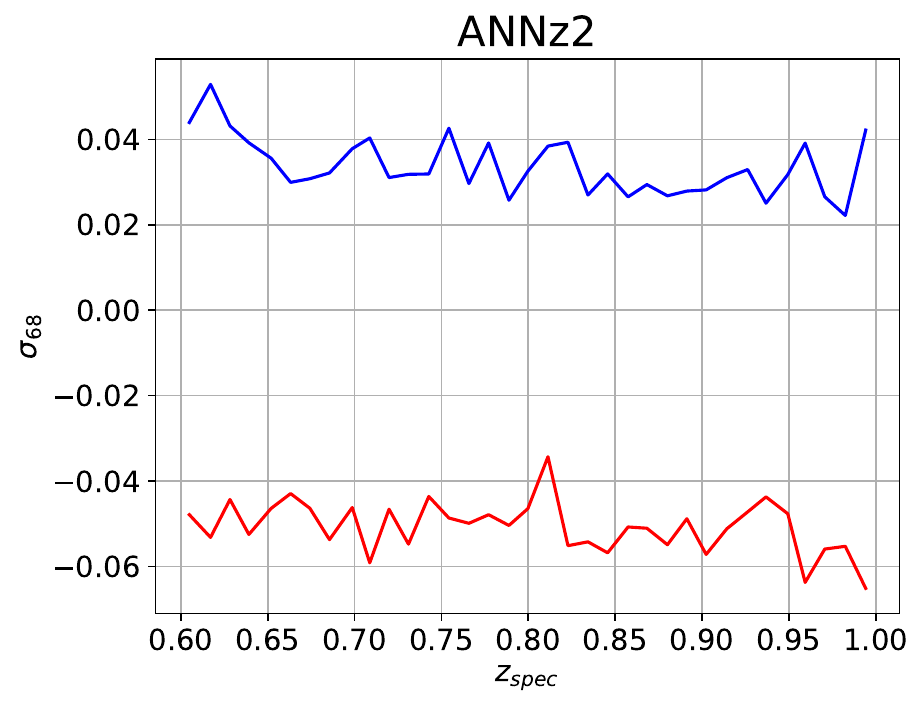}
        \caption{\texttt{ANNz2}}
        \label{fig:annz-sigma68}
    \end{subfigure}
    
    \vspace{0.5cm}
    
    \begin{subfigure}[b]{0.45\linewidth}
        \centering
        \includegraphics[width=\linewidth]{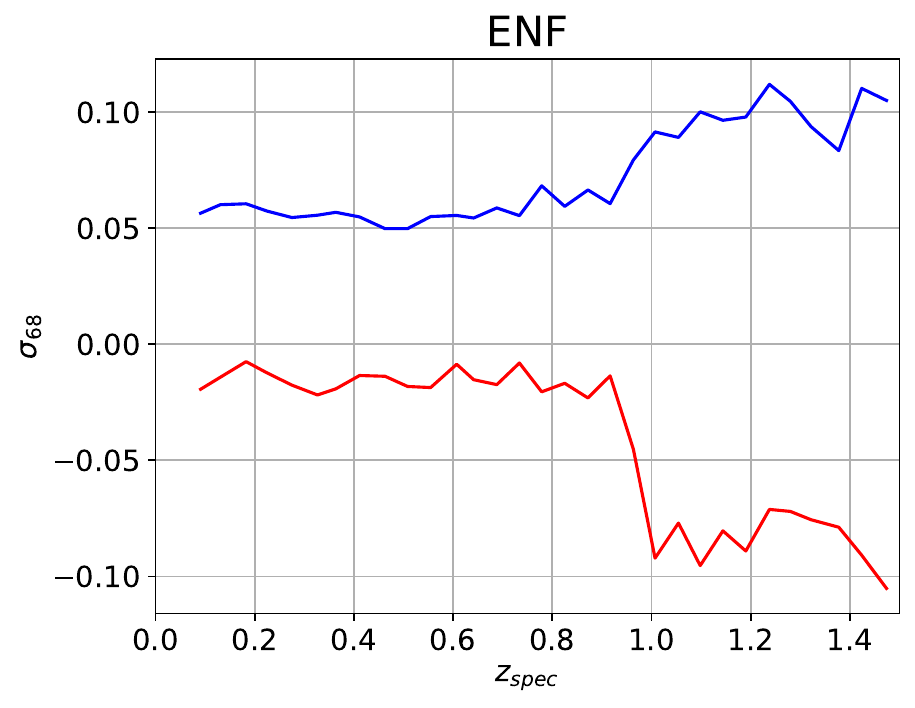}
        \caption{\texttt{ENF}}
        \label{fig:enf-sigma68}
    \end{subfigure}
    \hfill
    \begin{subfigure}[b]{0.45\linewidth}
        \centering
        \includegraphics[width=\linewidth]{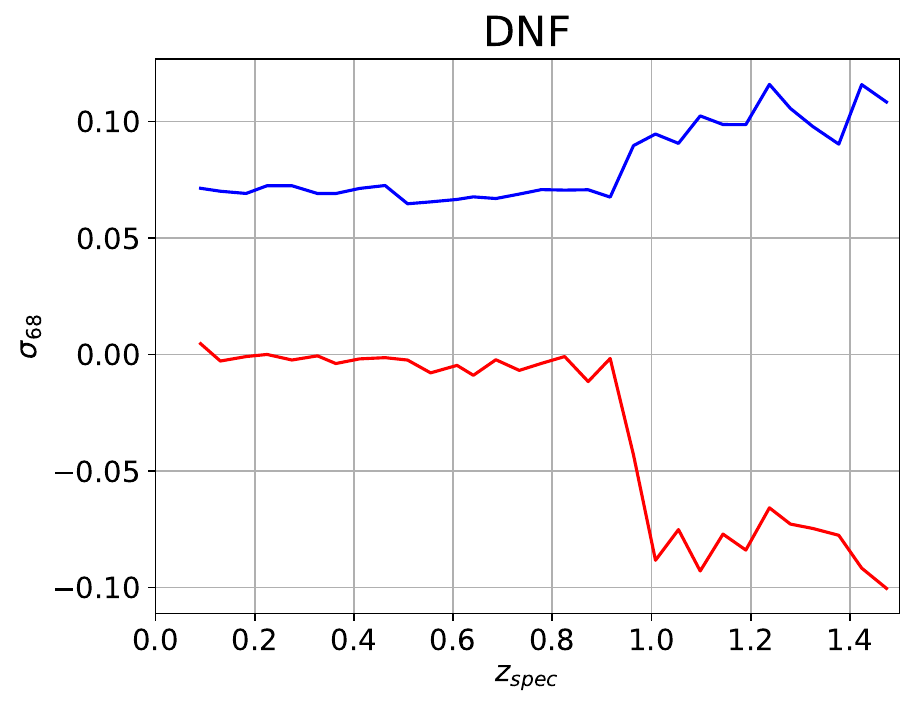}
        \caption{\texttt{DNF}}
        \label{fig:dnf-sigma68}
    \end{subfigure}
    
    \caption{$\sigma^{68}$ vs $z_{spec}$ bin. \textit{Blue}: positive error. \textit{Red}: negative error.}
    \label{fig:sigma_spec_z}
\end{figure}

This is consistent with Figures \ref{fig:anz2disp}, \ref{fig:bpz_vipers}, \ref{fig:enf2disp}, and \ref{fig:dnf2disp}. The asymmetry of overestimation by \texttt{ENF/DNF} and underestimation by \texttt{BPZ}, while \texttt{ANNz2} is nearly uniform for all redshift values.

We know there are certain galaxies that are not well estimated giving catastrophic redshift. Here, we choose $\sigma_z>\sigma_{99}$ as a catastrophic result. In table \ref{tab:tabsig} we compare the numbers of galaxies with catastrophic results, \texttt{ANNz2} has the best performance compared to the other estimators. Then, we compared the PDF performance in the catastrophic results, \texttt{BPZ} distribution modes $z_{mode}$ match exactly the photo-z result, but the other algorithms show discrepancy, with \texttt{ANNz2} as with the worst result.

\begin{table}
    \centering
    \begin{tabular}{c|c|c}
        Estimator & \# of galaxies with $\sigma_z>\sigma_{99}$ & \# of galaxies with $|z_{mode}-z_{phot}|>0.1$  \\
        \hline
        \texttt{ANNz2} & 146 & 81 \\
        \texttt{BPZ} & 1,283 &  0\\
        \texttt{ENF} & 1,179 & 38\\
        \texttt{DNF} & 1,084 & 27\\
        \hline
    \end{tabular}
    \caption{Catastrophic results w.r.t the PDF mode for each algorithm.}
    \label{tab:tabsig}
\end{table}

We compared the relative frequency of galaxies classified with Gaussian PDFs from the testing sample for each algorithm. In Figure \ref{fig:ngausssigma}, we show the number of Gaussians ($N_{Gauss}$) up to $|\sigma_z|$ compared to the number of galaxies in the test sample up to $|\sigma_z|$. \texttt{BPZ} has nearly 70\% Gaussian PDFs with bias bigger than 0.075 and \texttt{ANNz2} has nearly 60\%. \texttt{ENF} reaches a maximum of 25\% of Gaussian PDFs for very small bias, after $|\sigma_z|>0.025$, the frequency saturates and only 15\% of the PDFs are nearly Gaussian. The opposite happens to \texttt{DNF}, it is an improvement compared to \texttt{ENF}, but saturation is below 25\%. This result shows an important situation that may occur; even though the PDFs are smooth, there are catastrophic redshift results.

\begin{figure}
    \centering
    \includegraphics[width=.7\linewidth]{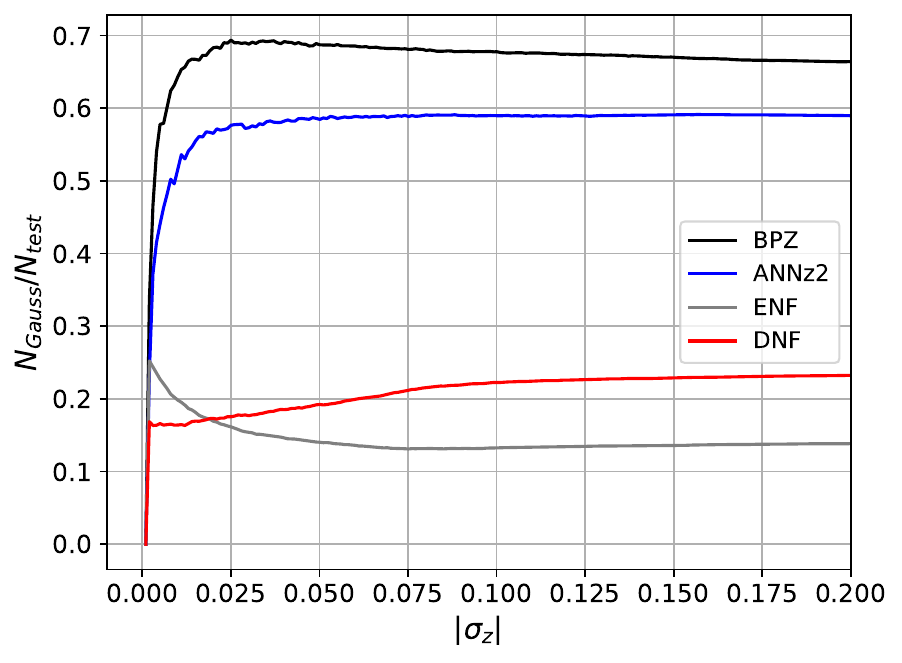}
    \caption{Relative frequency of Gaussian distributions compared to the total number of testing samples w.r.t. the absolute bias ($\sigma_z$) for each estimator. \texttt{BPZ}: black, \texttt{ANNz2}: blue, \texttt{ENF}: gray, and \texttt{DNF}: red.}
    \label{fig:ngausssigma}
\end{figure}

\subsection{Filters with respect to the bias}

 Another important aspect is how $\sigma_{68}$ behaves for each filter. We chose to analyse the behaviour for magnitudes greater than 20. First, we compare the bias normalized by $\sigma_{68}$ in Figure~\ref{fig:bias_filters}. The first panel shows the distribution for the filter $g$, because \texttt{DNF} has the larger error, the distribution (in red) is narrower than the other estimators. \texttt{ANNz2} and \texttt{BPZ} are centred at zero. \texttt{DNF} and \texttt{ENF} are centred at positive bias. The same pattern is seen in the second top panel for the filter $r$. The filters $i$ and $z$ show a different pattern, \texttt{ANNz2} is still centred at zero, while \texttt{BPZ} is centred at negative bias and \texttt{DNF} at positive bias. \texttt{ENF}'s filter $i$ is less positive than the other filters.
 Finally for filter $z$, in the last panel, the only difference from the previous one is \texttt{BPZ}'s distribution which is now more negative.

\begin{figure}
    \centering
    \includegraphics[width=\textwidth]{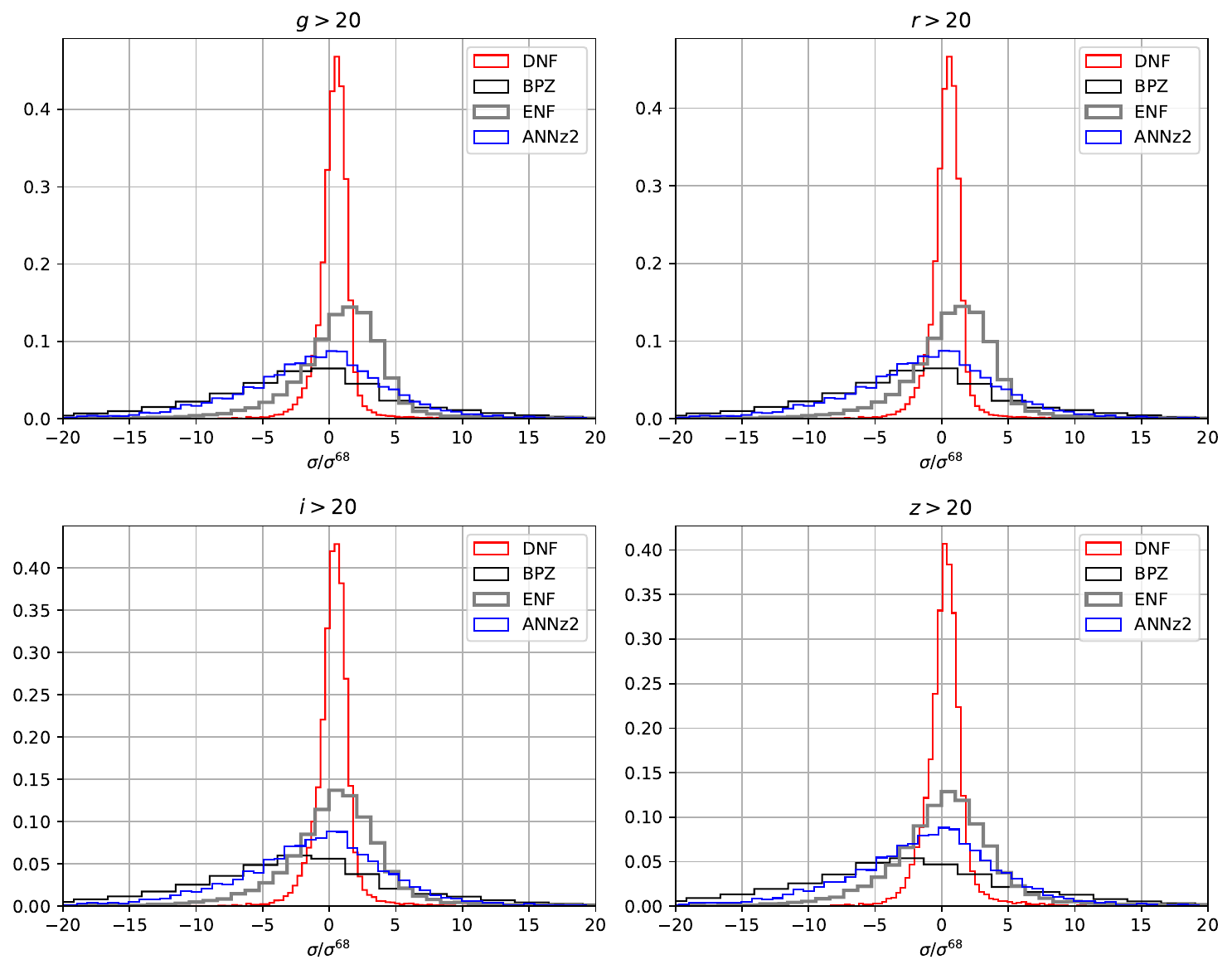}
    \caption{Redshift ias distribution for each filter normalized by $\sigma^{68}$. \textit{Red}: \texttt{DNF}, \textit{black}: \texttt{BPZ}, \textit{gray}: \texttt{ENF} and \textit{blue}: \texttt{ANNz2}.}
    \label{fig:bias_filters}
\end{figure}

Therefore, the best estimator is \texttt{ANNz2}, its error is consistent for all redshifts. The other estimators are less precise for higher redshift. For LSS analysis this is problematic, the ideal scenario should include a wider range of redshifts.

\subsection{Colour cut}

In the previous test, we based our $\sigma_{68}$ analysis w.r.t. the colour coverage of our spectroscopic sample, in Figure~\ref{fig:filter-spec}. For each estimator, we cut the redshifts restriction to the colours $1 < g-i < 4$ and $0.5 < r-i < 1.4$. Then, we selected the redshifts whose $\sigma_{68}$ is smaller than $10\%$ of the photo-z value. In table \ref{tab:tab_color}, there are the $z_{\rm eff}$ considering the colour cut and it is clear that they are compatible with the entire sample photo-z estimation. lastly, the number of objects reduces significantly to \texttt{DNF}, so depending on the use of the catalog, one must understand that an accurate representation of the filters by the training sample may reduce statistics. 

\begin{table}
    \centering
    \begin{tabular}{c|c|c}
        Estimator & $z_{\rm eff}$ & number of galaxies\\
        \hline
        \texttt{ANNz2} & 0.86 & 7,031,738\\
        \texttt{BPZ} & 0.84 & 7,012,076\\
        \texttt{ENF} & 0.82 & 6,304,106\\
        \texttt{DNF} & 0.84 & 4,727,115\\
    \end{tabular}
    \caption{Result after the colour cut.}
    \label{tab:tab_color}
\end{table}

We checked the bias distribution normalised by $\sigma_{68}$ in terms of the faintest part of each filter. This is shown in Figure~\ref{fig:bias_filters}, the only algorithm that is symmetric on the bias is \texttt{ANNz2}, \texttt{BPZ} underestimates the faint galaxies, while \texttt{DNF} and \texttt{ENF} overestimate them.

Fortunately, the criteria we have established for probability density functions have shown to be highly effective in eliminating suboptimal spectroscopic sample representations based on colour. This is evidenced in Figure~\ref{fig:color_panel} where we plot the $68\%$ confidence level of color surface distribution, which shows a notable reduction in subrepresented samples once we exclude the PDFs associated with high noise levels. The decrease in such subrepresentation becomes even more pronounced when we apply the Gaussian cut criterion, as depicted in Fig.~\ref{fig:color_panel2}. Within this Gaussian framework, the algorithm \texttt{ANNz2} emerges as the most accurate among the methods evaluated, whereas \texttt{ENF} experiences a substantial reduction, although this comes at the cost of some statistical accuracy. Ultimately, our careful selection strategy guarantees a reliable representation with respect to colour metrics, ensuring the overall robustness of our analytical approach.

In addition to the comparison of \texttt{DNF} from DES Y3 and our \texttt{DNF}, we repeat figure~\ref{fig:dnf_compared} but just with the galaxies lying inside the 1$\sigma$ region of the spectroscopic sample. The result can be seen in figure~\ref{fig:dnf_des_colors}, the dashed line represents the 1$\sigma$ region around the photo-z distributions. The cut significantly reduces the amount of objects to compare and we can see a peak difference between the two results around $z_p\sim 0.66$. Apart from this, the number of galaxies concentrate inside the 1$\sigma$ region. From this distribution, we found that $\langle \sigma \rangle =\langle z_{Pz \,\,cats}-z_{DESY3} \rangle= - 0.004$, similar to the previous result.

\section{Summary}\label{sec:conclusion}

This study explored the properties of four different photo-z estimators: \texttt{ANNz2}, \texttt{BPZ}, \texttt{ENF}, \texttt{DNF} with the same spectroscopic sample to train and/or test their precision. 

We selected 25,760 galaxies from four different spectroscopic surveys and cross-matched them with the DES Y3 BAO sample. These galaxies served to understand the redshift bias $\sigma$ and its 68th percentile $\sigma_{68}$ and are representative in terms of magnitude. We found that within a range of $0.79<z_p<0.85$ there is the lowest $\sigma$ for all the estimators we analysed. \texttt{DNF} has the biggest absolute value of the bias, while \texttt{ENF}, \texttt{ANNz2} and \texttt{BPZ} lose precision for a redshift range below 0.7 and greater than 0.9.

Computing the statistical metrics of the redshfit errors we found that the smallest bias belong to \texttt{BPZ} and the worst bias belongs to \texttt{DNF}. The least biased bin is $z_p\sim 0.8$ for all estimators, while the most biased bins are below 0.7 and above 0.9.

We also investigated the $\sigma_{68}$ metric, a robust metric to estimate the error of the algorithms. \texttt{ANNz2} and \texttt{BPZ} showed consistent $\sigma_{68}$ for all bins, while \texttt{ENF} and \texttt{DNF} have an asymmetric error distribution, with lower negative error values than positive ones and a considerable error increase for redshift greater than 0.9. As a consistent test, we compared with the DES Collaboration, because our and their results diverge slightly for higher redshift, we conclude that this algorithm is sensitive to the training set available especially in the high redshift end.

In another evaluation we conducted, our objective was to identify the optimal galaxies based on their redshift Probability Density Functions (PDFs). This endeavour resulted in a subset where \texttt{BPZ} identified the highest count of objects. Nevertheless, the Bayesian methodology intrinsic to \texttt{BPZ} tends to produce more homogenised PDFs. Conversely, \texttt{ANNz2}, although possessing a smaller subset than \texttt{BPZ}, demonstrated superior accuracy. On the other hand, \texttt{ENF} generated the least effective selection, associated with Gaussian PDFs, leaving a minimal quantity of galaxies suitable for a Large-Scale Structure (LSS) analysis, only 14 galaxies per degree squared. \texttt{DNF}, while being the least precise of the methods, still retained an adequate number of galaxies for cosmological investigations. Within the Small Peaks sub-sample, \texttt{ANNz2} outshone the other systems. Given the stringent criteria of the Gaussian subsample, it is less favourable for analyses requiring robust statistical outcomes. A noticeable effect of selecting PDFs is an improved representation in terms of colours compared with the spectroscopic sample; this finding underscores that such selection is significantly influenced by the magnitude space of the sample.

In case one wants to pick the best galaxies by removing the bins with the worst bias, they will find that \texttt{ANNz2} is the most robust algorithm for all criteria. \texttt{DNF} and \texttt{ENF} have a significant shift from zero. We also showed that even though the PDFs are smooth, there are catastrophic redshift results.

Finally, we conducted an analysis of how bias manifests in the context of faint galaxies. Our findings corroborate the previous analyses, indicating that \texttt{ANNz2} maintains consistency across all tested redshift ranges, whereas the other estimators exhibit tendencies to either overestimate or underestimate the photometric redshifts. Ultimately, among the four algorithms evaluated, \texttt{ANNz2} demonstrated success across the majority of the metrics we discussed, with comparatively fewer catastrophic failures. In an upcoming study, we aim to investigate whether this level of success exerts an influence on cosmological outcomes.

\section{Code availability}
The particular software packages used in this work will be accessible at \url{https://github.com/psilvaf/bao_pz} and \url{https://github.com/psilvaf/cat_org}. 

\acknowledgments

We acknowledge the use of the computational resources of the joint CHE / Milliways cluster, supported by a FAPERJ grant E26/210.130/2023. 

The authors would like to thank the anonymous referee who provided useful and detailed comments.

PSF thanks Brazilian funding agency CNPq for PhD scholarship GD 140580/2021-2. RRRR thanks CNPq for partial financial support (grant no. $309868/2021-1$).

The authors thank Juan De Vicente (Centro Investigaciones Energéticas, Medioambientales y Tecnológicas) for providing the \texttt{DNF} estimator code.

This project used public archival data from the Dark Energy Survey (DES). Funding for the DES Projects has been provided by the U.S. Department of Energy, the U.S. National Science Foundation, the Ministry of Science and Education of Spain, the Science and Technology FacilitiesCouncil of the United Kingdom, the Higher Education Funding Council for England, the National Center for Supercomputing Applications at the University of Illinois at Urbana-Champaign, the Kavli Institute of Cosmological Physics at the University of Chicago, the Center for Cosmology and Astro-Particle Physics at the Ohio State University, the Mitchell Institute for Fundamental Physics and Astronomy at Texas A\&M University, Financiadora de Estudos e Projetos, Funda{\c c}{\~a}o Carlos Chagas Filho de Amparo {\`a} Pesquisa do Estado do Rio de Janeiro, Conselho Nacional de Desenvolvimento Cient{\'i}fico e Tecnol{\'o}gico and the Minist{\'e}rio da Ci{\^e}ncia, Tecnologia e Inova{\c c}{\~a}o, the Deutsche Forschungsgemeinschaft, and the Collaborating Institutions in the Dark Energy Survey.
The Collaborating Institutions are Argonne National Laboratory, the University of California at Santa Cruz, the University of Cambridge, Centro de Investigaciones Energ{\'e}ticas, Medioambientales y Tecnol{\'o}gicas-Madrid, the University of Chicago, University College London, the DES-Brazil Consortium, the University of Edinburgh, the Eidgen{\"o}ssische Technische Hochschule (ETH) Z{\"u}rich,  Fermi National Accelerator Laboratory, the University of Illinois at Urbana-Champaign, the Institut de Ci{\`e}ncies de l'Espai (IEEC/CSIC), the Institut de F{\'i}sica d'Altes Energies, Lawrence Berkeley National Laboratory, the Ludwig-Maximilians Universit{\"a}t M{\"u}nchen and the associated Excellence Cluster Universe, the University of Michigan, the National Optical Astronomy Observatory, the University of Nottingham, The Ohio State University, the OzDES Membership Consortium, the University of Pennsylvania, the University of Portsmouth, SLAC National Accelerator Laboratory, Stanford University, the University of Sussex, and Texas A\&M University.
Based in part on observations at Cerro Tololo Inter-American Observatory, National Optical Astronomy Observatory, which is operated by the Association of Universities for Research in Astronomy (AURA) under a cooperative agreement with the National Science Foundation.

\bibliographystyle{JHEP}
\bibliography{references.bib}

\providecommand{\href}[2]{#2}\begingroup\raggedright\begin{thebibliography}{10}

\bibitem{arnouts1999measuring}
S.~Arnouts, S.~Cristiani, L.~Moscardini, S.~Matarrese, F.~Lucchin, A.~Fontana et~al., \emph{Measuring and modelling the redshift evolution of clustering: the hubble deep field north}, {\emph{Monthly Notices of the Royal Astronomical Society} {\bfseries 310} (1999) 540}.

\bibitem{guzzo2014vimos}
L.~Guzzo, M.~Scodeggio, B.~Garilli, B.~Granett, A.~Fritz, U.~Abbas et~al., \emph{The vimos public extragalactic redshift survey (vipers)-an unprecedented view of galaxies and large-scale structure at 0.5< z< 1.2}, {\emph{Astronomy \& Astrophysics} {\bfseries 566} (2014) A108}.

\bibitem{benitez2000bayesian}
N.~Benitez, \emph{Bayesian photometric redshift estimation}, {\emph{The Astrophysical Journal} {\bfseries 536} (2000) 571}.

\bibitem{feldmann2006zurich}
R.~Feldmann, C.M.~Carollo, C.~Porciani, S.~Lilly, P.~Capak, Y.~Taniguchi et~al., \emph{The zurich extragalactic bayesian redshift analyzer and its first application: Cosmos}, {\emph{Monthly Notices of the Royal Astronomical Society} {\bfseries 372} (2006) 565}.

\bibitem{brammer2008eazy}
G.B.~Brammer, P.G.~van Dokkum and P.~Coppi, \emph{Eazy: a fast, public photometric redshift code}, {\emph{The Astrophysical Journal} {\bfseries 686} (2008) 1503}.

\bibitem{10.1093/mnras/stt574}
M.~Carrasco~Kind and R.J.~Brunner, \emph{Tpz: photometric redshift pdfs and ancillary information by using prediction trees and random forests}, \href{https://doi.org/10.1093/mnras/stt574}{\emph{Monthly Notices of the Royal Astronomical Society} {\bfseries 432} (2013) 1483} [\href{https://arxiv.org/abs/https://academic.oup.com/mnras/article-pdf/432/2/1483/18463634/stt574.pdf}{{\ttfamily https://academic.oup.com/mnras/article-pdf/432/2/1483/18463634/stt574.pdf}}].

\bibitem{sadeh2016annz2}
I.~Sadeh, F.B.~Abdalla and O.~Lahav, \emph{Annz2: photometric redshift and probability distribution function estimation using machine learning}, {\emph{Publications of the Astronomical Society of the Pacific} {\bfseries 128} (2016) 104502}.

\bibitem{almosallam2016gpz}
I.A.~Almosallam, M.J.~Jarvis and S.J.~Roberts, \emph{Gpz: non-stationary sparse gaussian processes for heteroscedastic uncertainty estimation in photometric redshifts}, {\emph{Monthly Notices of the Royal Astronomical Society} {\bfseries 462} (2016) 726}.

\bibitem{de2016dnf}
J.~De~Vicente, E.~S{\'a}nchez and I.~Sevilla-Noarbe, \emph{Dnf--galaxy photometric redshift by directional neighbourhood fitting}, {\emph{Monthly Notices of the Royal Astronomical Society} {\bfseries 459} (2016) 3078}.

\bibitem{padmanabhan2005calibrating}
N.~Padmanabhan, T.~Budav{\'a}ri, D.J.~Schlegel, T.~Bridges, J.~Brinkmann, R.~Cannon et~al., \emph{Calibrating photometric redshifts of luminous red galaxies}, {\emph{Monthly Notices of the Royal Astronomical Society} {\bfseries 359} (2005) 237}.

\bibitem{hoshino2015luminous}
H.~Hoshino, A.~Leauthaud, C.~Lackner, C.~Hikage, E.~Rozo, E.~Rykoff et~al., \emph{Luminous red galaxies in clusters: central occupation, spatial distributions and miscentring}, {\emph{Monthly Notices of the Royal Astronomical Society} {\bfseries 452} (2015) 998}.

\bibitem{eisenstein2001spectroscopic}
D.J.~Eisenstein, J.~Annis, J.E.~Gunn, A.S.~Szalay, A.J.~Connolly, R.~Nichol et~al., \emph{Spectroscopic target selection for the sloan digital sky survey: The luminous red galaxy sample}, {\emph{The Astronomical Journal} {\bfseries 122} (2001) 2267}.

\bibitem{mcmahon2012vista}
R.~McMahon, \emph{The vista hemisphere survey (vhs) science goals and status}, {\emph{Science from the Next Generation Imaging and Spectroscopic Surveys} {\bfseries 37} (2012) }.

\bibitem{banerji2015combining}
M.~Banerji, S.~Jouvel, H.~Lin, R.~McMahon, O.~Lahav, F.~Castander et~al., \emph{Combining dark energy survey science verification data with near-infrared data from the eso vista hemisphere survey}, {\emph{Monthly Notices of the Royal Astronomical Society} {\bfseries 446} (2015) 2523}.

\bibitem{sevilla2021dark}
I.~Sevilla-Noarbe, K.~Bechtol, M.C.~Kind, A.C.~Rosell, M.~Becker, A.~Drlica-Wagner et~al., \emph{Dark energy survey year 3 results: Photometric data set for cosmology}, {\emph{The Astrophysical Journal Supplement Series} {\bfseries 254} (2021) 24}.

\bibitem{sanchez2014photometric}
C.~S{\'a}nchez~Alonso, M.~Carrasco~Kind, H.~Lin, M.~Serra~Ricart, F.B.~Abdalla, A.~Amara et~al., \emph{Photometric redshift analysis in the dark energy survey science verification data}, {\emph{Monthly notices of the Royal Astronomical Society. Oxford. Vol. 445, no. 2 (Dec. 2014), p. 1482-1506} (2014) }.

\bibitem{san2024dark}
L.T.~San~Cipriano, J.~De~Vicente, I.~Sevilla-Noarbe, W.~Hartley, J.~Myles, A.~Amon et~al., \emph{Dark energy survey deep field photometric redshift performance and training incompleteness assessment}, {\emph{Astronomy \& Astrophysics} {\bfseries 686} (2024) A38}.

\bibitem{abbott2022dark}
T.~Abbott, M.~Aguena, S.~Allam, A.~Amon, F.~Andrade-Oliveira, J.~Asorey et~al., \emph{Dark energy survey year 3 results: A 2.7\% measurement of baryon acoustic oscillation distance scale at redshift 0.835}, {\emph{Physical Review D} {\bfseries 105} (2022) 043512}.

\bibitem{collister2004annz}
A.A.~Collister and O.~Lahav, \emph{Annz: estimating photometric redshifts using artificial neural networks}, {\emph{Publications of the Astronomical Society of the Pacific} {\bfseries 116} (2004) 345}.

\bibitem{scodeggio2018vimos}
M.~Scodeggio, L.~Guzzo, B.~Garilli, B.~Granett, M.~Bolzonella, S.~De~La~Torre et~al., \emph{The vimos public extragalactic redshift survey (vipers)-full spectroscopic data and auxiliary information release (pdr-2)}, {\emph{Astronomy \& Astrophysics} {\bfseries 609} (2018) A84}.

\bibitem{newman2013deep2}
J.A.~Newman, M.C.~Cooper, M.~Davis, S.~Faber, A.L.~Coil, P.~Guhathakurta et~al., \emph{The deep2 galaxy redshift survey: Design, observations, data reduction, and redshifts}, {\emph{The Astrophysical Journal Supplement Series} {\bfseries 208} (2013) 5}.

\bibitem{le2005vimos}
O.~Le~Fèvre, G.~Vettolani, B.~Garilli, L.~Tresse, D.~Bottini, V.~Le~Brun et~al., \emph{The vimos vlt deep survey-first epoch vvds-deep survey: 11 564 spectra with 17.5$\leq$ i $\leq$ 24, and the redshift distribution over 0$\leq$ z$\leq$ 5}, {\emph{Astronomy \& Astrophysics} {\bfseries 439} (2005) 845}.

\bibitem{wang2020clustering}
Y.~Wang, G.-B.~Zhao, C.~Zhao, O.H.~Philcox, S.~Alam, A.~Tamone et~al., \emph{The clustering of the sdss-iv extended baryon oscillation spectroscopic survey dr16 luminous red galaxy and emission-line galaxy samples: cosmic distance and structure growth measurements using multiple tracers in configuration space}, {\emph{Monthly Notices of the Royal Astronomical Society} {\bfseries 498} (2020) 3470}.

\bibitem{alam2015eleventh}
S.~Alam, F.D.~Albareti, C.A.~Prieto, F.~Anders, S.F.~Anderson, T.~Anderton et~al., \emph{The eleventh and twelfth data releases of the sloan digital sky survey: final data from sdss-iii}, {\emph{The Astrophysical Journal Supplement Series} {\bfseries 219} (2015) 12}.

\bibitem{ross2017optimized}
A.J.~Ross, N.~Banik, S.~Avila, W.J.~Percival, S.~Dodelson, J.~Garcia-Bellido et~al., \emph{Optimized clustering estimators for bao measurements accounting for significant redshift uncertainty}, {\emph{Monthly Notices of the Royal Astronomical Society} {\bfseries 472} (2017) 4456}.

\bibitem{margoniner2008photometric}
V.~Margoniner and D.~Wittman, \emph{Photometric redshifts and signal-to-noise ratios}, {\emph{The Astrophysical Journal} {\bfseries 679} (2008) 31}.

\bibitem{reis2012sloan}
R.R.R.~Reis, M.~Soares-Santos, J.~Annis, S.~Dodelson, J.~Hao, D.~Johnston et~al., \emph{The sloan digital sky survey co-add: A galaxy photometric redshift catalog}, {\emph{The Astrophysical Journal} {\bfseries 747} (2012) 59}.

\bibitem{lima2008estimating}
M.~Lima, C.E.~Cunha, H.~Oyaizu, J.~Frieman, H.~Lin and E.S.~Sheldon, \emph{Estimating the redshift distribution of photometric galaxy samples}, {\emph{Monthly Notices of the Royal Astronomical Society} {\bfseries 390} (2008) 118}.

\bibitem{cunha2009estimating}
C.E.~Cunha, M.~Lima, H.~Oyaizu, J.~Frieman and H.~Lin, \emph{Estimating the redshift distribution of photometric galaxy samples--ii. applications and tests of a new method}, {\emph{Monthly Notices of the Royal Astronomical Society} {\bfseries 396} (2009) 2379}.

\bibitem{Ferreira:2025yvb}
P.S.~Ferreira and R.R.R.~Reis, \emph{{Influence of photometric galaxies redshift distribution in BAO estimation}},  \href{https://arxiv.org/abs/2501.04960}{{\ttfamily 2501.04960}}.

\bibitem{karim2025desi}
M.A.~Karim, J.~Aguilar, S.~Ahlen, S.~Alam, L.~Allen, C.A.~Prieto et~al., \emph{Desi dr2 results ii: Measurements of baryon acoustic oscillations and cosmological constraints}, {\emph{arXiv preprint arXiv:2503.14738} (2025) }.

\end{thebibliography}\endgroup

\end{document}